\documentclass[%
aps,prb,superscriptaddress,twocolumn,floatfix,10pt,longbibliography]{revtex4-2}
\usepackage[utf8]{inputenc}
\usepackage{amsmath,amssymb}
\usepackage{empheq}
\usepackage{float}
\usepackage{lipsum}
\usepackage{slashed}
\usepackage{mathtools,cuted}
\usepackage{esvect}
\usepackage[none]{hyphenat}
\usepackage{multirow}
\usepackage{xcolor}
\usepackage{soul}
\usepackage[export]{adjustbox}
\usepackage{graphicx}
\usepackage{dcolumn}
\usepackage{bm}
\usepackage{hyperref}
\usepackage{physics}
\usepackage{comment} 
\usepackage[mathlines]{lineno}
\usepackage[capitalise]{cleveref}
\usepackage{bbm}
\usepackage{xparse}

\newcommand{\A}{{\mathcal A}}
\newcommand{\G}{{\mathcal G}}
\newcommand{\ii}{\mathrm{i}}
\providecommand\i{}
\renewcommand{\i}{\mathrm{i}}
\newcommand{\ie}{\emph{i.e.},~}
\newcommand{\eg}{\emph{e.g.},~}
\newcommand{\sbar}{{\bar{s}}}

\newcommand{\FIXME}[1]{\textcolor{red}{FIXME: #1}}

\NewDocumentCommand{\thetaOne}{ O{} m m }{%
  \vartheta_1\IfValueT{#1}{^{#1}}\!\left( #2 \middle| #3 \right)
}

\newcommand{\elliptic}[4]{\vartheta\begin{bmatrix} #1 \\ #2 \end{bmatrix}\left( #3 \middle | #4\right)}

\begin{document}

\title{Hall viscosity and putative quantum Hall states without positive-definite K-matrix}

\author{E. Di Salvo}
\affiliation{Institute for Theoretical Physics, Utrecht University, 3584CC Utrecht, The Netherlands.}
\affiliation{Institute for Theoretical Condensed Matter Physics, Karlsruhe Institute of Technology, 76131 Karlsruhe, Germany}
\affiliation{Institute for Quantum Materials and Technologies, Karlsruhe Institute of Technology, 76344 Eggenstein-Leopoldshafen, Germany}

\author{D. Schuricht}
\affiliation{Institute for Theoretical Physics, Utrecht University, 3584CC Utrecht, The Netherlands.}

\author{J. K. Slingerland}
\affiliation{Department of Physics, Maynooth University, Ireland}

\author{M. Fremling}
\affiliation{Institute for Theoretical Physics, Utrecht University, 3584CC Utrecht, The Netherlands.}

\date{\today}

\begin{abstract}
We investigate putative quantum Hall effect states, labeled by their K-matrix equal to (1 1 3), by defining them on the torus and computing their Hall viscosity.
Such states have been introduced on the sphere as a phase distinct from Pfaffian and anti-Pfaffian ones.
This was done in order to explain certain results on thermal Hall conductivity in favor of particle-hole symmetric Pfaffian topological order in presence of Landau level mixing.
The requirements of boundary conditions, modular invariance and ground state degeneracy are enough to uniquely fix the form of the proposed wave functions.
We generalize a method to enforce them which we call monodromy matching and check our results on wave functions and Hall viscosity against realizations on the torus of Laughlin and hierarchical states.
We highlight the issues in the realization of these states, which turn out to exhibit the formation of clusters.
We show that the effect of anti-symmetrization on the system is not enough to prevent clustering; we compute the Hall viscosity for the Halperin version of these states and the fully anti-symmetrized one and we find them being dependent on the geometry and the particle number.
\end{abstract}

\maketitle{}

\section{Introduction}
\label{Introduction}
The advent of Topological Quantum Computing~\cite{k03} has revived the interest in anyonic excitations in fractional Quantum Hall effect (fQHE)~\cite{nssfd08}.
In particular, robustness and fault-tolerance provided by the topological protection are crucial features in order to develop Quantum Computing at its full potential.
Among the zoo of fQHE states observed through the years, $\nu=5/2$ stands at the forefront for its rich exotic properties including non-Abelian braiding statistics and Majorana edge modes.
Although its experimental discovery traces back to 1987, its full theoretical understanding is still lacking today.
To describe the ground state of this system, multiple wave functions representing different topological states have been proposed, among which we list the Pfaffian by Moore and Read~\cite{mr91}, the anti-Pfaffian~\cite{lhr07,lrnf07} and their particle-hole-symmetric version~\cite{s15}.
A later approach was advanced in terms of the so-called $A$-phase in~\cite{ddm23}, where an anomalous topological $\nu=5/2$ quantized fQHE phase was identified by the means of exact diagonalization of the Coulomb Hamiltonian in the second Landau level at an intermediate strength of the Landau level mixing.
The proposal implies a trial ground state wave function in the spherical geometry which is a fully anti-symmetrized 113 Halperin wave function. 
The regular Halperin wave function (\ie not anti-symmetrized) is known to be clustering when off-diagonal elements are larger then diagonal ones, hence it breaks the screening hypothesis in the plasma analogy~\cite{qjm89}.
Some debate in the community as whether anti-symmetrization is enough to prevent clustering~\cite{s23,ddm23reply} emerged after the publication of~\cite{ddm23}.
The authors proposed the hypothesis of anti-symmetrization being a mechanism to prevent clustering, but it has been numerically shown on the sphere that such putative fQHE states still do cluster~\cite{sb24}.

We propose to add to the debate by analyzing the Hall viscosity of such states, which for topologically ordered states has a finite value~\cite{asz95}.
In order to retrieve it, we study the problem of putting the 113 state on a toroidal geometry.
The nontrivial boundary conditions on the torus enforce more restrictions over the form of proposed wave functions; however, such strictness can be turned into an effective tool since it enables us to fix uniquely their form.
Historically, the first to deal with this problem for fQHE has been Haldane, who described Laughlin states on a torus and highlighted the remarkable principle that the two-particle short-distance behavior of wave functions has to be independent from the geometry~\cite{hr85}.
Starting from the same principle, we can explicitly write down the Jastrow factors in all the cases we analyze, which will be the starting points of our construction.
Originally, the definition on the torus of integer quantum Hall effect (iQHE) by Niu, Thouless and Wu provided an elegant formulation of the problem of the quantization of electric conductivity~\cite{ntw85}.
Moreover, in more recent years the description of quantum Hall effect (QHE) states on compact manifolds has deemed necessary to understand the topological order that characterize them.
It manifests itself in the non-trivial dependence of the structure of the Hilbert space from the genus of the surface $g$, as for the ground state degeneracy, equal to $|\det K|^g$~\cite{wn90}.

The striking richness and complexity of physical phenomena in fQHE has pushed an intense effort to describe them all in a unifying framework.
The most straightforward approach consists in writing down the $N$-particles wave functions of the ground state at a certain filling fraction $\nu \equiv N/N_\phi = p/q$ where $p$ and $q$ are integers: apart from the ground-breaking work by Laughlin~\cite{l83}, two methods have paved the way.
One, pioneered by Moore and Read, maps conformal field theory (CFT) correlators into ground state wave functions~\cite{mr91}; the other is the composite fermion approach, introduced by Jain, that exploits the concept of flux attachment by Laughlin~\cite{jk97}.
Both of them fail however to provide a viable ground state wave function for 113 state (see Sec~\ref{SEC:Failure} for details).
On the other hand, long distance behavior of fQHE states can be analyzed by the means of Chern-Simons field theories: Wen and Zee provided a full classification of Abelian fQHE states in terms of integer-valued matrices, labeled K-matrices, by computing the shift in the flux number $N_\phi$ of all these theories when described on a sphere~\cite{wz92}, which is a well-defined topological quantity.
Even though these field theories fully capture the topological properties of the ground states, it do not cover the short distance and dynamical properties of the system.

The computation of the Hall viscosity in our case is performed numerically using a Monte Carlo algorithm to sample the overlap between states after slight deformations of the toric geometry.
Another route to compute viscosity has been pursued in recent years of writing the fQHE states as matrix product states exploiting CFT's state-operator correspondence, pioneered by Zaletel and Mong~\cite{zaletel2013}.
As a starting point, however, the knowledge of a parent Hamiltonian of the considered states is needed~\cite{sfjj18,kssj23,bons20}.
This is not accessible for states as 113.

Our work is structured as follows: in section \ref{sec:QH-wfn-torus} we introduce the problem of defining a valid fQHE wave function on the torus, starting from the basic example of Landau levels and we provide some general statements regarding periodicity and modularity of wave functions.
In section \ref{SEC:Failure} we show why wave functions for states with non-positive definite K-matrix cannot be analytically constructed by the means of the CFT correspondence.
Section \ref{sec:Monodromy_matching} introduces monodromy matching that generalize the methods developed in the previous one to encompass fQHE states with non-positive definite K-matrix and in section \ref{sec:V-Realizations} specific realizations for different models can be found.
Section \ref{sec:Hall viscosity} finally provides the values of Hall viscosity for all the previously mentioned states.
In the Appendices are stored all the definitions and derivations needed for the main text.

\section{Quantum Hall wave function on the torus}
\label{sec:QH-wfn-torus}

In this section we give an overview on how to construct fQHE wave functions on the torus.
The first realization of such a program is due to Haldane~\cite{hr85} for Laughlin's wave functions with filling fraction $\nu=1/q$ (odd integer values of $q$ corresponds to states formed by condensates of fermionic excitations, even ones to bosonic).
Even in this simpler case, many features of fQHE are evident: the relation among ground state degeneracy, manifold genus and center-of-mass (CoM) zeroes, the action of CoM piece as an extra particle with charge $-qe$ and the geometrical relation with quantized conductivity via the Niu-Thouless-Wu formula~\cite{ntw85,Fradkin91}.

\begin{figure}
    \centering
    \includegraphics[width=\columnwidth]{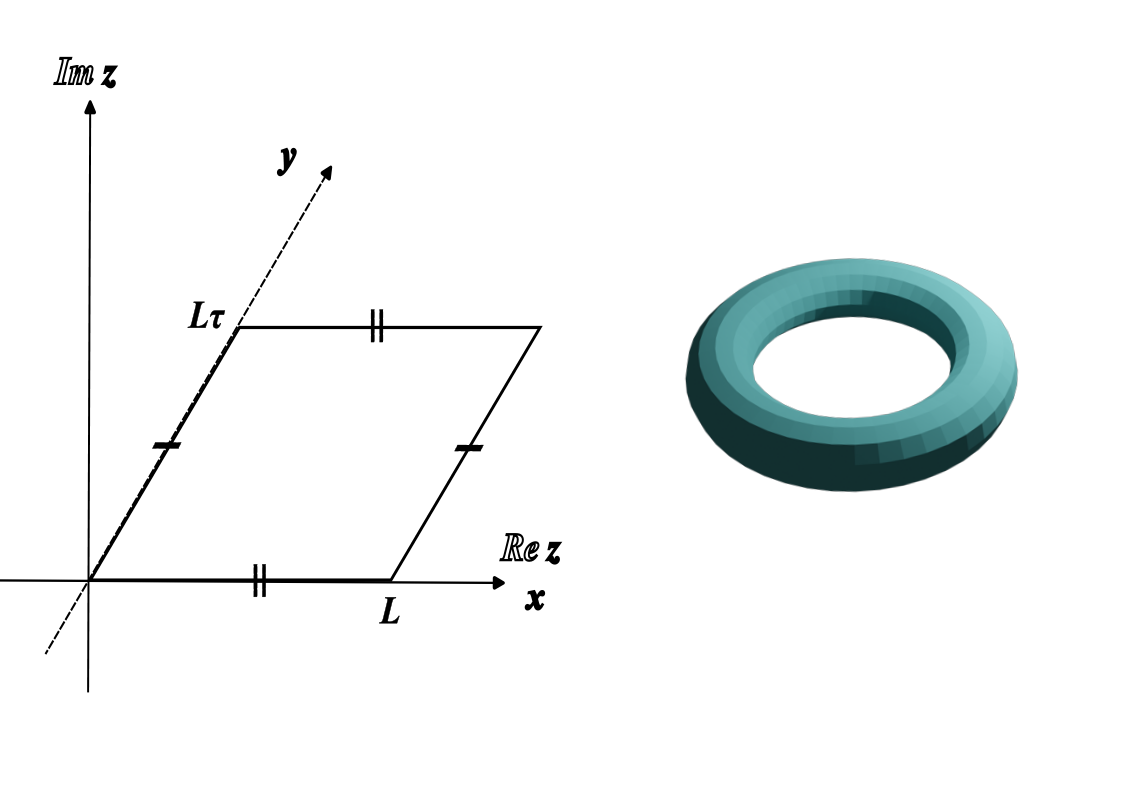}
    \caption{Torus representation on the two-dimensional complex plane and in its embedding in three dimensions. While in the first case it amounts to the quotient space obtained after the identification of opposite sides of the unit cell, in the second case the periodicity is imposed from the beginning on the surface of the ``doughnut". On the complex plane we identify the coordinates along the real axis and the straight line generated by the complex number $\tau$ as $x$ and $y$.}
    \label{Fig:torus}
\end{figure}

We consider a torus $\mathbb{C}/(L\mathbb{Z}+L\tau\mathbb{Z})$ of complex characteristic $\tau = \tau_1 + \ii\tau_2$ and area $A=L^2\tau_2$.
On the torus itself we fix the particle coordinates to be $z_j = L(x_j+\tau y_j)$.
The setup is also depicted in Figure~\ref{Fig:torus}.
The $(x,y)$ components are reduced coordinates that have real values between $0$ and $1$.
However such a construction is not unique: equivalent tori can be obtained either by shifting the characteristic by an integer $T:\tau\to\tau + n$ or by reparametrizing $S:\tau\to-1/\tau$, $z\to(\tau/|\tau|) z$.
This redundance goes under the name of modular invariance and all the observables and wave functions must transform in a covariant way, \ie they must only take into account the change of coordinates due to reparametrization. 
The name modular transformation comes from the fact that all transformations are elements of the modular group $SL(2,\mathbb{Z})$, the action of which preserves both periodicity and orientation of the torus.

Finally, the torus can be described as a quotient of the complex plane under the identifications $z\to z+L$ and $z\to z+L\tau$ and we are employing this description throughout our paper.
This is equivalent to imposing generalized periodic boundary conditions to the wave function, as allowed by gauge invariance, and usual periodic boundary conditions to the observables~\cite{Fradkin91}.

\subsection{Landau levels on the torus}
\label{Landau Levels on the torus}
In the presence of a magnetic field $B$, the quantum mechanical problem of the motion of an electron (or any charged particle) is drastically different from the one of a free particle: in the latter case, particles move freely and they are just determined by their momentum $\vec{p}$, describing a spectrum of continuum excitations, while in the former the spectrum is quantized and highly degenerate~\cite{l30}.
The Hamiltonian reads
\begin{equation}
    \label{LLHamiltonian}
    H = \frac{1}{2m}\vec{\Pi}\cdot\vec{\Pi} = \hbar\omega_B\left(a^\dagger a + \frac{1}{2}\right)~,
\end{equation}
where $\omega_B = eB/m$ is the cyclotron frequency and $m$ is the electronic mass; the introduction of $\omega_B$ automatically implies the presence of a magnetic length $l_B = \hbar(eB)^{-1/2}$.
The ladder operators $a^\dagger$ and $a$ are those that move particles between different energy levels, which are the Landau levels.
Both $a^\dagger$ and $a$ are defined in terms of the mechanical momentum vector $\vec{\Pi} = \vec{p} + e\vec{A}$, where $\vec{A}$ is the vector potential generated by the magnetic field $\vec{B}=\vec{\nabla}\times\vec{A}$, as
\begin{equation}
    \label{LadderOpA}
    a = \frac{1}{\sqrt{e\hbar B}}(\Pi_x - \ii\Pi_y)~.
\end{equation}
The mechanical momentum cannot be canonically transformed into the generalized ones, $\vec{p}$, and this explains the different nature of free particle's spectrum in presence or absence of magnetic field.

\begin{figure}[!t]
    \centering
    \includegraphics[width=\columnwidth]{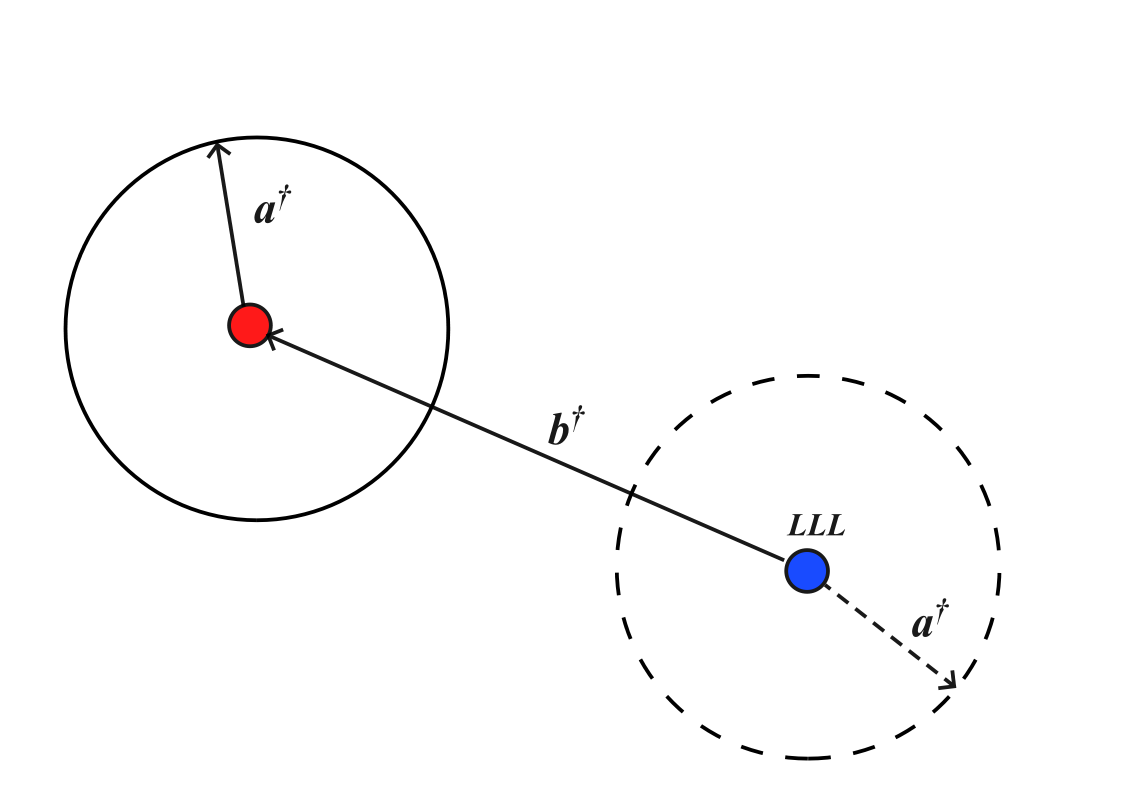}
    \caption{Action of creation operators $a^\dagger$ and $b^\dagger$ on the semiclassical trajectories of electrons. The first moves between different Landau levels, which measures the cyclotron radius. The second span the orbital degeneracy in the energetic spectrum, which moves the guiding center.}
    \label{Fig:LLs}
\end{figure}

It is straightforward to identify the lowest Landau level (LLL) as the states that are annihilated by the operator $a$.
Any representative of higher excited states labelled by the quantum number $n$ is obtained by successive applications of the creation operator $a^\dagger$.
To cycle through all states within a Landau level, one introduces the operator
\begin{equation}
    \label{LadderOpB}
    b = \frac{1}{\sqrt{e\hbar B}}(\tilde\Pi_x - \ii\tilde\Pi_y)~,
\end{equation}
and its complex conjugate $b^\dagger$.
Here the new set of guiding center coordinate momenta is $\vec{\tilde\Pi} = \vec{p} - e\vec{A}$.
The action of these operators is depicted in Figure \ref{Fig:LLs}.

To obtain specific expressions for wave functions one has to pick a gauge.
On the torus, a convenient choice is the $\tau$-gauge, but one can also exploit another common realization, which is the symmetric gauge~\cite{Fradkin91}.
Using $\tau$-gauge, in reduced coordinates $(x,y)$ corresponds the choice of the vector potential to be
\begin{equation}
    \label{taugauge}
    \vec{A}_\tau = 2\pi N_\phi By \begin{pmatrix} 1 \\ 0 \end{pmatrix},
\end{equation}
corresponding to a magnetic field $\vec{B} = B\vec{z}$ pointing in the out-of-plane direction.
With this gauge choice, any LLL one-particle states can be written as
\begin{equation}
    \label{tauLLL}
    \phi_{LLL,\tau}(z,\bar{z}) = e^{\ii\pi\tau N_\phi y^2}f(z)~,
\end{equation}
where $f(z)$ stands for any holomorphic function with generalized periodic boundary conditions on the torus~\cite{Fradkin91}.
The non-analytic phase in front goes under the name Gaussian piece.


We will be employing the $\tau$-gauge for most of the computations.
Gauge dependence will be suppressed from the notation, unless necessary to highlight a different gauge choice.
Another remark on notation must be done: we are considering the torus with length $L=1$ throughout the rest of the paper without any loss of generality.

\subsection{Magnetic translations and the Heisenberg group}
\label{Magnetic translations and the Heisenberg group}
The absence of mechanical momentum as a good quantum number of the magnetic system does not only reflect on the spectrum, but also on its translation properties: the momentum operator $\vec{p}$ generates translations, but the Hamiltonian eigenstates are not eigenstates of $\vec{p}$.
On the other hand, we can define new translation operators, a.k.a. magnetic translation operators $t(z)$~\cite{z64}, that are diagonal with respect to the eigenstates of the Hamiltonian.
The magnetic translation operators fulfill the Girvin-MacDonald-Platzman algebra~\cite{gmp85,gmp85_2}
\begin{equation}
    \label{GMP}
    t(z)t(z') = t(z+z')e^{\ii\tau_2 A(z,z')/(2l_B^2)}~,
\end{equation}
in which the term $A(z,z')$ is the area spanned by the parallelogram generated by the two vectors $z$ and $z'$ represented by complex numbers.
In the $\tau$-gauge, the two elementary translation operators on the torus can be written as~\cite{fhs14}
\begin{eqnarray}
    \label{tOp}
    &t_1 = t(\epsilon) = e^{\epsilon\partial_x} ~,\\
    &t_2 = t(\tau\epsilon) = e^{\epsilon\partial_y + 2\ii\pi x}~\nonumber.
\end{eqnarray}
The elementary translation is given by the shift of the coordinates by the finite quantity $\epsilon = 1/N_\phi$ and it respects the characteristic equation for the Heisenberg group
\begin{equation}
    \label{HeisenbergGroup}
    t_1t_2 = e^{2\ii\pi\epsilon}t_2t_1~,
\end{equation}
hence the wave functions must be representations of the latter.

For many-body wave functions containing $N$ different particles, translations can be defined as single-particle translations $t^{(j)}_k(z)$, which are equivalent to \eqref{tOp} accompanied by the label $j$ of the particle they act on.
Moreover, from the elementary translations operators \eqref{tOp}, we can construct CoM translations~\cite{fhs14}, defined as
\begin{equation}
    \label{eq:T-CoM}
    T_k = \prod_{j=1}^N t^{(j)}_k~,
\end{equation}
where $k=\{1,2\}$.
They also belong to Heisenberg group, but in this case the phase acquired after a shift is simply $e^{2\ii\pi N\epsilon} = e^{2\ii\pi\nu}$.
It is then straightforward to show that $T_1$ and $T_2^q$ commute between themselves and any translationally invariant many-body Hamiltonian.
Given that the fundamental representation of \eqref{eq:T-CoM} has dimension $q$ on the torus, there are at least $q$ different degenerate copies of the ground state; for non-Abelian states there will be a multiple of $q$ states.

\subsection{Periodic boundary conditions and modular transformations}
\label{Periodic boundary conditions and modular transformations}
Putting an electronic system on a geometry with periodic boundary conditions is not a trivial task;
in the specific case of QHE, boundary conditions are intimately related to the transverse Hall conductivity.
In particular, a non-zero Hall conductivity implies that a LLL wave function's phase cannot be fully determined on the space of different possible realizations of the boundary conditions themselves~\cite{Fradkin91}.
In other words, imposing periodic boundary conditions as
\begin{equation}
    \label{PBC}
    \left[t^{(j)}_k\right]^{N_\phi}\psi(z_1,\dots,z_N) = e^{2\ii\pi\varphi_k}\psi(z_1,\dots,z_N)~,
\end{equation}
not only reduces the initial Heisenberg group to the subgroup $SU(N_\phi)$~\cite{h11}, but it also cannot be uniquely defined for any value of $\varphi_k$.
In fact, in the presence of a non-vanishing Hall conductivity, the phase of the wave function acquires an extra term when the boundary phase is sent $\varphi_k\to\varphi_k+2\pi$; it then cannot be universally defined in the space of different boundary phases $\varphi_k$.
This is the key ingredient to obtain a non-vanishing first Chern number~\cite{k85}.
A warning is also necessary: when we consider expressions like \eqref{PBC} for QHE states, gauge phases developed by the translation of a single particle are not considered, since they are eventually eliminated by a gauge transformation, and all the results are stated after the change of gauge has been applied.
That's why we refer to this boundary conditions for wave functions as generalized ones.

Translations of the CoM coordinates are also possible and any QHE wave function has to be an eigenfunction of the two operators that generate them, \ie $T_1$ and $T_2^q$.
The related eigenvalue can be understood as the momentum of the CoM in the two directions.
As a matter of fact, the CoM component can be viewed as a particle moving in a specific direction in order to balance all the relative momenta from the single components~\cite{hr85}.

We can now explicitly impose boundary conditions on a fQHE wave function with filling fraction $\nu$ of the following form
\begin{eqnarray}
    \label{QHEwavefunction}
    &&\psi_\nu(z_1,\dots,z_N|\tau) = ~\\
    &&\qquad\mathcal{N}_N(\tau)G(y_1,\dots,y_N|\tau)\A\left \{f(z_1,\dots,z_N)\right\}~, \nonumber
\end{eqnarray}
where the letter $\A$ indicates the anti-symmetrization of the arguments, since the wave function is defined from electronic degrees of freedom.
By $G(y_1,\dots,y_N)$ we consider the many-body version of the Gaussian piece in \eqref{tauLLL}, dictated by gauge choice (the so-called $\tau$-gauge in our case)~\cite{fhs14}
\begin{equation}
    \label{GaussianPiece}
    G(y_1,\dots,y_N|\tau) = e^{\ii\pi\tau N_\phi\sum_{j=1}^N y^2_j}~.
\end{equation} 
When anti-symmetrization is considered, one can split the generic function $f$ into the relative and CoM coordinates dependent pieces
\begin{equation}
    \A\left \{f(z_1,\dots,z_N)\right\} = \A\left \{J(z_1,\dots,z_N|\tau)\mathcal{F}(Z|\tau)\right\}~,
\end{equation}
that in the literature are usually labelled as Jastrow and CoM pieces.
The capital $Z=\sum_{j=1}^Nz_j$ is the CoM coordinate and when more particle species are present it is represented as a vector with a number of entries equal to the number of particle species.
The Jastrow part and CoM piece have to be adapted according to the physics of the state considered and we can show how the first piece can be obtained by some general considerations.
This problem reduces to finding a CoM piece for any state defined by a set of topological properties, encoded in the K-matrix.

Starting from the Gaussian piece \eqref{GaussianPiece}, it transforms as
\begin{equation}
    \label{GaussianPBC}
    G(\dots,y_j + 1, \dots|\tau) = G(\dots,y_j, \dots|\tau) e^{\ii\pi\tau N_\phi(2y_j + 1)}~,
\end{equation}
after translating the first particle in the $\tau$ direction of a length $L$, while it is invariant under translations in the direction of the real axis.
The same relation applies for any other particle labeled by the coordinate $y_j$.
The phase appearing in \eqref{GaussianPBC} is crucial because it allows the definition of the Jastrow and CoM parts in terms of quasi-periodic holomorphic functions in the $\tau$ direction.
Such functions are exactly theta functions~\cite{Mumford07}, see Appendix~\ref{App: Theta functions} for definitions and more details.
In addition to this, the normalization is constructed to be dependent only from the particle number $N$, the torus characteristic $\tau$~\cite{f17} and the K-matrix.

Before delving into the details of the explicit construction, we are briefly mentioning the effect of reparametrizations of the torus on the wave function.
For the wave function to be invariant under the action of the modular group, the reparametrized one has to be equivalent in the Hilbert space to the original one, \ie it has to be mapped into the previous one by a gauge transformation.
Again, we analyze the action of the modular group on the Gaussian piece
\begin{eqnarray}
    \label{GaussianModTrans}
    &&T:~G(y_1,\dots,y_N|\tau+1) = \nonumber\\
    &&\qquad\qquad G(y_1,\dots,y_N|\tau)e^{\ii\pi N_\phi\sum_{j=1}^N y^2_j}~,\\
    &&S:~G(y_1',\dots,y_N'|-1/\tau) = \nonumber\\
    &&\qquad\qquad G(y_1,\dots,y_N|\tau)e^{-\ii\pi(N_\phi/\tau)\sum_{j=1}^Nz_i^2},
\end{eqnarray}
where the new coordinates $y_i'=\Im{\frac{\tau}{|\tau|}z_i}$ are defined according to the $S$-transformation rules.
Then all the other pieces must transform under $S$- or $T$-transformations in a way that cancels the pieces in the previous expression that are not constant phases but depend on $z$ and $\bar{z}$~\cite{fhs14}.

We conclude that writing the Jastrow part and CoM piece as theta function is not enough. 
The normalization has to transform accordingly, since it depends on the torus characteristic itself, and it can be expressed in terms of Dedekind eta functions $\eta(\tau)$ and a power of the imaginary component $\tau_2$~\cite{f17}.
It has been shown, however, that this is just an approximation of the dependence of normalization on torus geometry.
Other relevant contributions are expected for instance when the degenerate limit $\tau\to0$ is approached.

A fundamental tool to describe analytically QHE states on the torus is given by theta functions $\vartheta(z|\tau)$.
Their definition and properties in multiple dimensions are contained in the Appendix \ref{App: Theta functions}.
Throughout the paper, the notation is going to articulate as follows: when the pedix is $1$, like $\vartheta_1(z)$, we consider mono-dimensional theta functions with both characteristics of $1/2$ (note that $\vartheta_1(z)\sim z$ for $z\to0$); when they are specified, they are usually written in the matrix form $\begin{bmatrix} \vec{\alpha} \\ \vec{\beta}\end{bmatrix}$; when both the pedix and the matrix notation is missing, we consider mono-dimensional theta functions with arbitrary characteristics.

\section{CFT correspondence and K-matrices}
\label{SEC:Failure}
In this section we try and construct the wave function for a fQHE state with non-positive K-matrix and thereby show why it fails.
The starting point is the correspondence between CFT correlators and fQHE wave functions: any polynomial part of the wave function \eqref{QHEwavefunction} for $N$ particles, \ie the product between Jastrow factors and CoM piece, can be expressed as the $N$-point correlation function of some primary operators of a CFT~\cite{mr91,hhsv17}.
The choice of the CFT is not arbitrary and bears information regarding the edge states of the theory~\cite{w92}.
Moreover, when $N$-particle ground states with different species with particle number $N^{(\alpha)}$ are considered, it is customary to consider the magnetic fluxes $N_\phi^{(\alpha)}$ for each of the species.
The connection between K-matrix and Chern-Simons theory is contained in Appendix~\ref{App: Effective Field Theory of Quantum Hall effect}.

\subsection{Positive-definite K-matrices}

As a starting point, we briefly summarize how the CFT construction works in the case of positive K-matrices. 
The simplest way to set up the CFT is to use the Coulomb gas formalism for compactified bosons on the torus~\cite{hhsv17}.
In this formalism, any correlation function of vertex operators $V_{\vec{q}} = e^{\ii\vec{q}\cdot\vec{\phi}(z,\bar{z})}$ can be mapped into a problem of classical charged particles interacting via a Coulomb-like potential: in particular, the expectation function vanishes when the neutrality condition is not met.
Such construction can be extended by considering a smeared charged background, which allows for different neutrality conditions and central charges, via the introduction of an appropriate charged background operator $Q_{\mathrm{bg}}$~\cite{DiFrancesco:1997nk}.
The $N$-point function in the CFT then is given by the expectation value
\begin{equation}
    \label{CFTNpointfunc}
    \langle V_{\vec{q}_1}(z_1,\bar{z}_1)\dots V_{\vec{q}_N}(z_N,\bar{z}_N) Q_{\mathrm{bg}}\rangle~,
\end{equation}
of $N$ different vertex operators placed on each particle's position and labelled by their charge vector $\vec{q}$.
This procedure allows the imposition of the correct boundary conditions, and the background charge is responsible for the appearance of the Gaussian piece in \eqref{QHEwavefunction}~\cite{mr91,fhs14}.
The wave function can then be cast into the following general fashion by employing charge vectors~\cite{fhs14}
\begin{eqnarray}
    \label{CFTwavefunction}
    &&\psi_\nu(z_1,\dots,z_N|\tau) = \mathcal{N}_N(\tau)G(y_1,\dots,y_N|\tau)\\
    &&\qquad\times\prod_{i\neq j}^N\thetaOne{z_i-z_j}{\tau}^{\vec{q}_i\cdot\vec{q}_j} \mathcal{F}_{\vec{h}}(Z|\tau)~, \nonumber
\end{eqnarray}
where
\begin{equation}
    \label{eq:CFTCoM}
    \mathcal{F}_{\vec{h}}(Z|\tau) = \sum_{\vec{q}\in\Gamma}e^{\ii\pi\tau(\vec{q}+\vec{h})^2}e^{2\ii\pi(\vec{q}+\vec{h})\cdot\sum_{i=1}^N\vec{q}_i z_i}~.
\end{equation}
The lattice of charge vectors, $\Gamma$, is defined on a $d$-dimensional space, where $d$ is the number of species and is spanned by the primary vectors $\vec{q}_\alpha$, defined by the relation
\begin{equation}
    \label{eq:ChargeVectKMatr}
    K_{\alpha\beta} = \vec{q}_\alpha\cdot\vec{q}_\beta~.
\end{equation}
Here, the indices $\alpha$ and $\beta$ label particle species and not particle number.
The vector $\vec{h}$ implements the correct boundary conditions and properties of the zeroes of the CoM piece.

It is possible to rewrite \eqref{eq:CFTCoM} as a multidimensional theta function~\cite{Mumford07} in the form of
\begin{equation}
  \label{eq:CFTwfn-Kmatrix}
  \mathcal{F}_{\vec{s}}(\vec Z|\tau) = \elliptic{K^{-1}\vec s}{\vec 0}{K\vec Z }{\tau K},
\end{equation}
where we identify $s_{\alpha}=\vec{h}\cdot\vec{q}_{\alpha}$.
We introduce here the shorthand $N_\phi^{(\alpha)}/N^{(\alpha)}\equiv\nu_\alpha$, valid for the remainder of the work.
To discuss the action of the CoM operators $T_j$ on the CoM pieces, we will here define reduced operators $\tilde T_j$, which act only on $\mathcal{F}_{\vec{s}}(\vec Z|\tau)$, while keeping all the other terms unchanged after a translation of $\nu_\alpha$ on the torus.
They are defined through the relation 
\begin{equation}
  \label{eq:CoM-Tilde-Def}
 T_j\left\{\psi(\vec{z})\right\}  = G(\vec y) J(\vec z) \times \tilde T_j  \left\{\mathcal{F}(\vec Z)\right\}~,
\end{equation}
where one can show that 
\begin{align}
  \label{eq:CoM-Tilde}
\tilde T_1  = T_1 &= e^{\sum_\alpha 
\nu_\alpha \frac{\partial}{\partial {X_\alpha}}}\nonumber\\
\tilde T^c_2  &= e^{\i2\pi c Z}e^{\i\pi\tau\nu c^2} e^{c\sum_\alpha \nu_\alpha \frac{\partial}{\partial {Y_\alpha}}}.
\end{align}
We keep the arbitrary (real) power, $c$, of $\tilde T_2$ here to show that there is a non-linear relation in the $\tau$-dependent part of the definition.

With the action of $\tilde T_1$ and $\tilde T_2$ defined in \eqref{eq:CoM-Tilde}, it is easy to show that they act on $\mathcal{F}_{\vec{s}}(Z|\tau)$ is as
\begin{align}
  \tilde T_1\mathcal{F}_{\vec{s}}(\vec Z|\tau)&=e^{2\i\pi\sum_\alpha 
s_\alpha\nu_\alpha}\mathcal{F}_{\vec{s}}(\vec Z|\tau)\nonumber\\
    \tilde T_2\mathcal{F}_{\vec{s}}(\vec Z|\tau)&=\mathcal{F}_{\vec{s}+\vec{1}}(\vec Z|\tau).
\end{align}
The equations above show that $\mathcal{F}_{\vec{s}}$ is an eigenstate of $T_1$, whereas $T_2$ cycles through a subset of the states.

\subsection{Non-positive definite K-matrices}

We now assume that we have a non-positive K-matrix and consider the effects of such assumption on the CFT approach.
The first problem that arises is that the charge vectors in \eqref{eq:ChargeVectKMatr} cannot be constructed anymore, at least not as long as the scalar product is a positive-definite quadratic form.
This can be illustrated in the 2$\times$2 case: here, a non-positive eigenvalue happens when $K_{12}^2>K_{11}K_{22}$. However one can show that for a positive-definite quadratic form then $|K_{12}|=|\vec q_1\cdot \vec q_2| \leq |\vec q_1||\vec q_2| = \sqrt{K_{11}K_{22}}$, which is precisely what is violated when one of the eigenvalues is negative.
We are then left with two options: either we define the wave function only in terms of the K-matrix elements, with no reference to charge vectors as is \eqref{eq:CFTwfn-Kmatrix}, or we find a way to employ a different, non-positive definite, quadratic form that maps charge vectors to K-matrix elements.

The first approach is bound to fail for the following reason: in the CoM piece in \eqref{eq:CFTwfn-Kmatrix}, the expression is only well defined if the sum in the theta function can actually be performed.
This sum contains the term $\exp{\i\pi\tau \vec l^T K \vec l}$ which converges exponentially fast in the summation index $\vec l$ when the K-matrix is positive definite, but which diverges exponentially when this is not the case (see Appendix \ref{App: Theta functions} for details).
The divergence can be explained by the fact that the K-matrix now contains negative eigenvalues, and they introduce terms that are exponentially growing in the sum.
Such negative eigenvalues also have a physical meaning, as they are related to counter-propagating edge modes in the finite-volume planar geometry.
It means that it is not possible to rewrite this term as a function of any K-matrix.
One may also wonder whether the $d$-dimensional theta function $\vartheta(K\vec{Z}|\tau K)$ can be "analytically continued" on the whole $\mathbb{R}^{d,d}$, but this is not possible without introducing divergent contributions to the sum or singularities in its domain~\cite{Mumford07}.
The second way is also forbidden by the fact that introducing a different quadratic form entails a redefinition of the metric $g^{\mu\nu}$ of the CFT Lagrangian $\mathcal{L} = (1/2)\times\int dx g^{\mu\nu}\partial_\mu\phi \partial_\nu\phi$.
A non-positive definite form implies a non-positive definite Lagrangian, which, in turn, makes the theory non-unitary.
It was shown by Read~\cite{r09} that the CFT correspondence breaks down when non-unitary CFTs are considered.
It is then not possible to define the CoM of such wave functions in the same fashion as for positive K-matrices, only in terms of multi-dimensional theta functions.

A result that can be borrowed from the CFT correspondence is the definition of the Jastrow factor in terms of K-matrix elements
\begin{eqnarray}
    \label{JastrowCFT}
    &&J\left(z^{(1)}_1,\dots,z^{(1)}_{N_1},z^{(2)}_1,\dots,z^{(n)}_{N_n}|\tau\right) = \nonumber\\
    &&\qquad\qquad\qquad\prod_{a,b}^n\prod_{i\neq j}^{N_a,N_b}\thetaOne{z^{(a)}_i-z^{(b)}_j}{\tau}^{K_{ab}},
\end{eqnarray}
since it has the expected short-distance scaling and braiding properties.
The K-matrix dictates quantitatively such repulsion among different and equal species~\cite{wz92} since it relates the magnetic fluxes induced by all the species with their particle number
\begin{equation}
    \label{Kmatrix}
    N_\phi^{(a)} = \sum_{b=1}^n K_{a,b}N_b~.
\end{equation}
By now, requiring periodic boundary conditions, \ie that equation \eqref{PBC} holds for all particle types $j$, leads to the following pseudo-periodicity condition on the CoM term
\begin{align}
  \label{CoMeqs}
  \mathcal{F}(\vec{Z}+\vec{e}_j|\tau) &= (-1)^{N_\phi-K_{jj}}\mathcal{F}(\vec{Z}|\tau)~,\\
  \mathcal{F}(\vec{Z}+\vec{e}_j\tau|\tau) & =\nonumber \\
  &(-1)^{N_\phi-K_{jj}}e^{-2\ii\pi \sum_{k}K_{jk}Z_{k}}e^{-\i\pi\tau K_{jj}}\mathcal{F}(\vec{Z}|\tau).\nonumber
\end{align}
In the above equation, the only restriction on the K-matrix is that it is integer-valued. In deriving the above relation, we made use of the relation $\sum_{i\neq j}K_{i,j}=N_\phi-K_{jj}$.
This shows explicitly that the unknown reminder in \eqref{QHEwavefunction}, \ie the CoM piece, is a function dependent only on the CoM coordinates $\vec{Z}$.
In principle one could add constant phases $e^{\ii\phi_j^{(1)}}$ and $e^{\ii\phi_j^{(2)}}$ to these expressions. However, we can, without loss of generality, set these to zero since CoM translation operators can always recover them.

We are now in the position to investigate this problem by finding CoM terms that are solutions to \eqref{CoMeqs}.

\section{Monodromy matching}
\label{sec:Monodromy_matching}
In this section, we propose an ansatz for the CoM piece of multi-component wave functions with a given K-matrix. To do so, we will generalize Haldane's original construction for the Laughlin state \cite{hr85} to arbitrary K-matrices.

Let's consider the generic case of a multi-component state's wave function determined by a given K-matrix \eqref{QHEwavefunction}.
Our main observation is that we can fix the form of $\mathcal{F}$ by considering solely its monodromy properties: by this, we mean the phase that is picked up as you traverse the ``edges" of the torus unit cell defined on the complex plane.
For positive-definite and holomorphic CoM pieces, this corresponds precisely to the number of zeroes of $\mathcal{F}$~\cite{Mumford07} and are already encoded in the K-matrix. Note, though, that one has to consider not only the zeroes of one's own species but also the zeroes of the other species, as we will explain below.
Using this as a starting point, we consider a plain holomorphic function
\begin{equation}
  \label{G_simple}
  \mathcal{G}(z) = \mathcal{G}_0(\tau)e^{\i 2\pi z \beta}\prod_{i=1}^N\thetaOne{z+\alpha_i}{\tau},
\end{equation}
that we factorize in a product of theta functions in terms of its zeroes.
If we send $z\to z+1$ and $z\to z+\tau$, we can fix boundary conditions, imposed by equation \eqref{CoMeqs}.
Using quasi-periodic properties of theta functions contained in Appendix~\ref{App: Quasi-periodic properties} we find that
\begin{align}
  \label{G_simple_trans}
  \mathcal{G}(z+1)  &= (-1)^Ne^{\i2\pi\beta} \mathcal{G}(z) ,\nonumber\\
  \mathcal{G}(z+\tau) & = (-1)^N e^{-2\i\pi\left( \tau \frac{N}2  - \tau\beta+N z+\sum_{i=1}^{N}\alpha_i\right)}\mathcal{G}(z).
\end{align}
We will now compare equation \eqref{CoMeqs} with  $\mathcal F$  and equation \eqref{G_simple_trans} with $\mathcal{G}$, where we make the identification $Z_j=z$.
In this way we can fix the coefficient $2\beta =_2 N_\phi$;  we use the $=_t$ as shorthand for this expression holding modulo $t$ in the real part.
Moreover, we obtain that $K_{jj} = N$, namely the number of zeroes for the $j$-particles in the CoM function is precisely dictated by the K-matrix value $K_{jj}$.
Finally, we can relate the coefficients $\alpha_j$ by taking the logarithm of the expressions.
If we consider $-\alpha_i$ as the i:th zero of the CoM piece, then we see that there is a family of states that satisfy the conditions.
If we, for instance, restrict ourselves to the one-dimensional case (\ie the family of Laughlin states) where $K=K_{11}=q$, we get
$\sum_{i=1}^{q}\alpha_i =_1 \beta+\tau\beta$, with $\beta = \frac{q N_e}2 + \mathbb Z$.
This shows explicitly how the $q$-dimensional set of states for a given Laughlin state is encoded in the constraint of the CoM zeroes $-\alpha_i$, as it was observed by Haldane in~\cite{Haldane1985Laughlin}.
Finally, one should also note that in this case all the zeroes $\alpha_i$ are independent of $Z$.

\subsection{Two-dimensional positive definite K-matrices}

The key challenge when considering higher-dimensional K-matrices is that there are now multiple variables $Z_j$, each with their own factorization in the form of (\ref{G_simple}), and they all have to be compatible with each other.
As a consequence, the sum over $\alpha$ will always be a linear combination of $Z_j$ with integer coefficients.
For instance, for the 331 state, these equations would reduce to 
\begin{align}
  \sum_{i=1}^3\alpha_i^{(1)} &=_1  K_{12}Z_{2} = Z_{2}\nonumber\\
  \sum_{i=1}^3\alpha_i^{(2)} &=_1  K_{21}Z_{1} = Z_{1}\nonumber
\end{align}
where for simplicity $\beta=0$.
A solution to this set of equations is 
\begin{equation}
  \label{G_simple331}
  \mathcal{G}_{331}(Z_1,Z_2) = \thetaOne[2]{Z_1}{\tau} \thetaOne{Z_1+Z_2}{\tau} \thetaOne[2]{Z_2}{\tau}
\end{equation}
where we can see that both $Z_1$ and $Z_2$ have three zeroes since $K_{11}=K_{22}=3$, but they only see one zero that depends on the position of the other variable since $K_{12}=1$.
The physical interpretation of this approach relates the total number of zeroes of a certain component of the CoM variable to the relative magnetic fluxes per particle attached to that species, \ie $K_{aa}$.
On the other hand, the number of zeroes that depend from other components of the CoM vector is given by the number of fluxes induced by the presence of other species.

The solutions for the 331 state are easily generalized to higher dimensional K-matrices, using the same intuition exposed above.
These considerations naturally lead to the following ansatz for $\mathcal{G}$
\begin{equation}
    \label{MMAnsatzGeneral}
    \mathcal{G}(\vec{Z}|\tau) = \prod_{\vec{i}\in I_n}\prod_{k=1}^{d_{\vec{i}}}\elliptic{\alpha_{\vec{i},k}}{\beta_{\vec{i},k}}{Z_{\vec{i}}}{\tau},
\end{equation}
where $I_n$ labels all possible (ordered) combinations of the numbers $1,\ldots,n$, and $k$ runs over the $d_{\vec{i}}$ different factors in the product.
Below, we will see that the $d_{\vec{i}}$ are determined by the K-matrix.
The coefficients $\alpha_{\vec{i},k}$ and $\beta_{\vec{i},k}$ are generalizations of the $\alpha_i$ and $\beta$ that was used in equation \eqref{G_simple331}, while $Z_{\vec{i}}=\sum_{k\in\vec i}Z_k$ is the sum of all the elements of $\vec{Z}$ included in $\vec{i}$.

In practice, we write the most general product of different one-dimensional theta functions and restrict the conditions on the parameters $k$, $\alpha_{\vec{i},k}$, and $\beta_{\vec{i},k}$.
The heuristic for the number of zeroes is that if $\mathcal{G}$ has a factor, say $\vartheta(Z_a+Z_b+\ldots)$, then the group $Z_a$ will see a zero for the group $Z_b$ and vice versa.
Their location will be determined by the $\alpha_{\vec{i},k}$ and $\beta_{\vec{i},k}$ coefficients.
This expression is not an eigenstate of the CoM translation operators.
As a consequence, a linear combination of it and its translations along the real and $\tau$-axis must be used to generate momentum eigenstates.

Repeating the calculations that led to \eqref{G_simple_trans}, now using the generalized theta functions (transformation \eqref{P1D} and \eqref{QP1D} of the Appendix) we find that $\mathcal{G}$ has the following transformation properties
\begin{align}
  \label{G_gernal_trans}
  \mathcal{G}(\vec{Z}+\vec{e}_j)  &= \prod_{\vec i \ni j}\prod_{k=1}^{d_{\vec{i}}}e^{2\ii\pi\alpha_{\vec{i},k}} \mathcal{G}(\vec{Z}) ,\\
  \mathcal{G}(\vec{Z}+\vec{e}_j\tau) & =  \prod_{\vec i \ni j}\prod_{k=1}^{d_{\vec{i}}}
  e^{-\ii\pi\tau}e^{-2\ii\pi(Z_{\vec i}+\beta_{\vec i,k})}\mathcal{G}(\vec{Z})\nonumber,
\end{align}
where the product $\prod_{\vec i \ni j}$ should be read as all ordered sets of numbers $1,\ldots,n$ that contain $j$.
Comparing \eqref{G_gernal_trans} with \eqref{CoMeqs} we find first that, for any value of $\tau$ and $\vec Z$,
\begin{equation}
\sum_{\vec i \ni j} d_{\vec i} = K_{jj}~,
\end{equation}
and secondly that 
\begin{eqnarray}
  \label{eq:Monodromy_constraints}
  K_{ab} &=& \sum_{\vec{i} \ni a,b }d_{\vec i}~.
\end{eqnarray}
Here $\sum_{\vec{i} \ni a,b }$ means the sum over all ordered vectors $\vec i$ that contain both $a$ and $b$.

For $2\times2$ matrices, the solution to \eqref{eq:Monodromy_constraints} is unique, but for a higher number of components, the decomposition is not; hence, in this work, we will mainly focus on the $2$-component case. The equation for $d$ is then
\begin{equation}
    K = \begin{pmatrix}
        d_1 + d_{12} & d_{12} \\
        d_{12} & d_2 + d_{12}
    \end{pmatrix} ~,
\end{equation}
that leads to $d_{12} =K_{12}$, $d_{1} =K_{11} - K_{12}$ and $d_{2} =K_{22} - K_{12}$.
In this case, \eqref{MMAnsatzGeneral} reduces to
\begin{eqnarray}
    \mathcal{G}_{\vec{\alpha},\vec{\beta}}(\vec{Z}|\tau) = \prod_{i=1}^n\prod_{k=1}^{d_n}\elliptic{\alpha_{i,k}}{\beta_{i,k}}{Z_i}\tau\nonumber \\
    \times\prod_{i_1<i_2=1}^n\prod_{k=1}^{d_{i_1i_2}}\elliptic{\alpha_{i_1i_2,k}}{\beta_{i_1i_2,k}}{Z_{i_1i_2}}{\tau}~.
\end{eqnarray}
Thus for two-dimensional positive definite K-matrices, it is always possible to construct a version of the center-of-mass piece using monodromy matching that reproduces the same properties of the multi-dimensional theta function.
We checked this explicitly for the case with K-matrix 331 in Appendix~\ref{App: Relation between different center-of-mass pieces}.

\subsection{Two-dimensional non-positive definite K-matrices}
\label{Non-positive definite K-matrices}

We now turn our attention to cases with non-positive definite K-matrices.
What effectively happens in this situation is that one or more of the $d_{\vec{i}}$ that solves \eqref{eq:Monodromy_constraints} becomes negative.
That $d_{\vec{i}}$ becomes negative is easily seen in the 2-dimensional case, since when $K_{12}>K_{11}$, then $d_1=K_{11}-K_{12}$ is negative. At the heuristic level, we can interpret a negative $d_{\vec{i}}$ as a pole in the wave function or, even more specifically, the difference between poles and zeros in the wave function.

We can make this statement more precise by modifying our ansatz in \eqref{MMAnsatzGeneral} to read
\begin{equation}
    \label{MMAnsatzGeneral_II}
    \mathcal{G}(\vec{Z}|\tau) = \prod_{\vec{i}\in I_n}\frac{\prod_{k=1}^{d_{\vec{i}}^U}\elliptic{\alpha^U_{\vec{i},k}}{\beta^U_{\vec{i},k}}{Z_{\vec{i}}}{\tau}}{\prod_{k=1}^{d_{\vec{i}}^L}\elliptic{\alpha^L_{\vec{i},k}}{\beta^L_{\vec{i},k}}{Z_{\vec{i}}}{\tau}},
    \end{equation}
such that it also includes $Z$-dependent terms in the denominator.
It leads to the more general constraint
\begin{eqnarray}
  \label{eq:Monodromy_constraints_II}
  K_{ab} &=& \sum_{\vec{i} \ni a,b }(d^U_{\vec i}-d^L_{\vec i}).
\end{eqnarray}
which only fixes the differences between $d^U_{\vec i}$ and $d^L_{\vec i}$ and not the sum.
As such, one is, in principle, free to add pairs of nominators and denominators at will.
One can however argue that a minimal realization of $\mathcal{G}(\vec{Z}|\tau)$ will have either $d^U_{\vec i}=0$ or $d^L_{\vec i}=0$ for a given $\vec i$.

We see from \eqref{eq:Monodromy_constraints_II} that poles will always be present in the CoM piece of a fQHE when the K-matrix is non-positive definite.
However, this is not the only way to meet the required boundary conditions.
By relaxing the requirement of a holomorphic CoM term we can obtain similar results.
The idea is to replace each pole $\thetaOne[-1]z\tau$ with the complex conjugate $\overline{\thetaOne{z}{\tau}}=-\thetaOne{-\bar z}{-\bar \tau}$, since both $\thetaOne[-1]z\tau$ and $\overline{\thetaOne{z}{\tau}}$ will pick up the same phase when encircled.

The non-holomorphic CoM piece will then read
\begin{equation}
    \label{MMAnsatzGeneral_III}
    \G(\vec{Z}|\tau) = \prod_{\vec{i}\in I_n}
    \prod_{k=1}^{d_{\vec{i}}^H}\elliptic{\alpha^H_{\vec{i},k}}{\beta^H_{\vec{i},k}}{Z_{\vec{i}}}{\tau}
    \prod_{k=1}^{d_{\vec{i}}^N}\elliptic{\alpha^N_{\vec{i},k}}{\beta^N_{\vec{i},k}}{-\bar Z_{\vec{i}}}{-\bar\tau},
\end{equation}
where we used the label (H)olomorphic and (N)on-holomorphic for the respective parts.
It is then simple to find the transformation properties after a translation to be
\begin{align}
  \label{G_gernal_trans_II}
  \G(\vec{Z}+\vec{e}_j)  &= \prod_{\vec i \ni j}
  \prod_{k=1}^{d^H_{\vec{i}}}e^{2\ii\pi\alpha^H_{\vec{i},k}}
  \prod_{k=1}^{d^N_{\vec{i}}}e^{-2\ii\pi\alpha^N_{\vec{i},k}} 
  \G(\vec{Z}),\nonumber\\
  \G(\vec{Z}+\vec{e}_j\tau) & =  \prod_{\vec i \ni j}
  \prod_{k=1}^{d^H_{\vec{i}}}e^{-\ii\pi\tau}e^{-2\ii\pi(Z_{\vec i}+\beta^H_{\vec i,k})}\\
  &\quad\times\prod_{k=1}^{d^N_{\vec{i}}}e^{\ii\pi\bar\tau}e^{-2\ii\pi(-\bar Z_{\vec i}-\beta^N_{\vec i,k})}
  \G(\vec{Z})\nonumber~.
\end{align}
Upon inspection it should be evident that the results in \eqref{G_gernal_trans_II} can never be made to fit the requirements in \eqref{CoMeqs} since the former contains terms of the form $\bar\tau$ and especially $\bar Z$, not present in \eqref{CoMeqs}.

The solution here is however straightforward. We note that since the number of zeros is matching, the problem is the incorrect scaling.
This can be accounted for by introducing an extra scale factor, which can be regarded of as a modification of the Gaussian piece.
By using that $\bar\tau=\tau -  2\ii\tau_2$ and $\bar z = z - 2\ii\tau_2 y$, we find that the pieces that we need to cancel will transform to produce the factors 
$ \exp{2\pi\tau_2\sum_{\vec i \ni j} d^N_{\vec{i}} }$
and 
$\exp{4\pi \tau_2 \sum_{\vec i \ni j} d^N_{\vec{i}} Y_{\vec i}}$.
If we define $M_{j,k}=\sum_{\vec i \ni j,k} d^N_{\vec{i}}$ as the non-holomorphic counterpart of \eqref{eq:Monodromy_constraints_II} we find that the extra scale factor that we need to insert is
\begin{equation}
  G_{AH}(\vec{Y}|\tau) = e^{-2\pi\tau_2\sum_{j,k}Y_jM_{jk} Y_k}~.
  \end{equation}
With this modification the CoM piece satisfies the same boundary conditions as the one containing poles.

We would at this point like to point out that monodromy matching, together with a non-holomorphic CoM is not the only way to obtain a valid wave function for a non-positive K-matrix.
Another construction developed by Girvin and Jach on the plane~\cite{gj84} and later extended to the torus~\cite{f13}, is to modify the Jastrow factor itself.
The construction is illustrated using the Laughlin state's Jastrow factor
\begin{eqnarray}
    \label{EQ:GJ-WF}
    J(\vec z|\tau) = \prod_{i\neq j}^N\thetaOne[q]{z_i-z_j}{\tau} \left|\thetaOne{z_i-z_j}{\tau}\right|^{2k},
\end{eqnarray}
where extra non-analytic pieces have been added to yield a stronger short-distance repulsion, controlled by $k>0$.
As only absolute values of $\thetaOne{z_i-z_j}{\tau}$ are introduced the total number of fluxes in the state is fixed and therefore the filling fraction of the state stays unchanged.
Using the techniques of Ref.~\cite{fhs14}, the construction above can be generalized to higher dimensional K-matrices in a straightforward manner~\cite{hhsv17}\footnote{M. Fremling, private communication}.
It is worth pointing put that these constructions contain non-holomorphic components in the Jastrow factors, and thus scales with $N_e$.
It is therefore expected that their non-LLL component is significantly higher than the construction above in equation \eqref{MMAnsatzGeneral_III}, which has a constant non-holomorphic contribution.
In both cases, we will be dealing with wave functions that are not completely in the LLL.

\begin{figure}[!t]
  \centering
  \includegraphics[width=0.9\columnwidth]{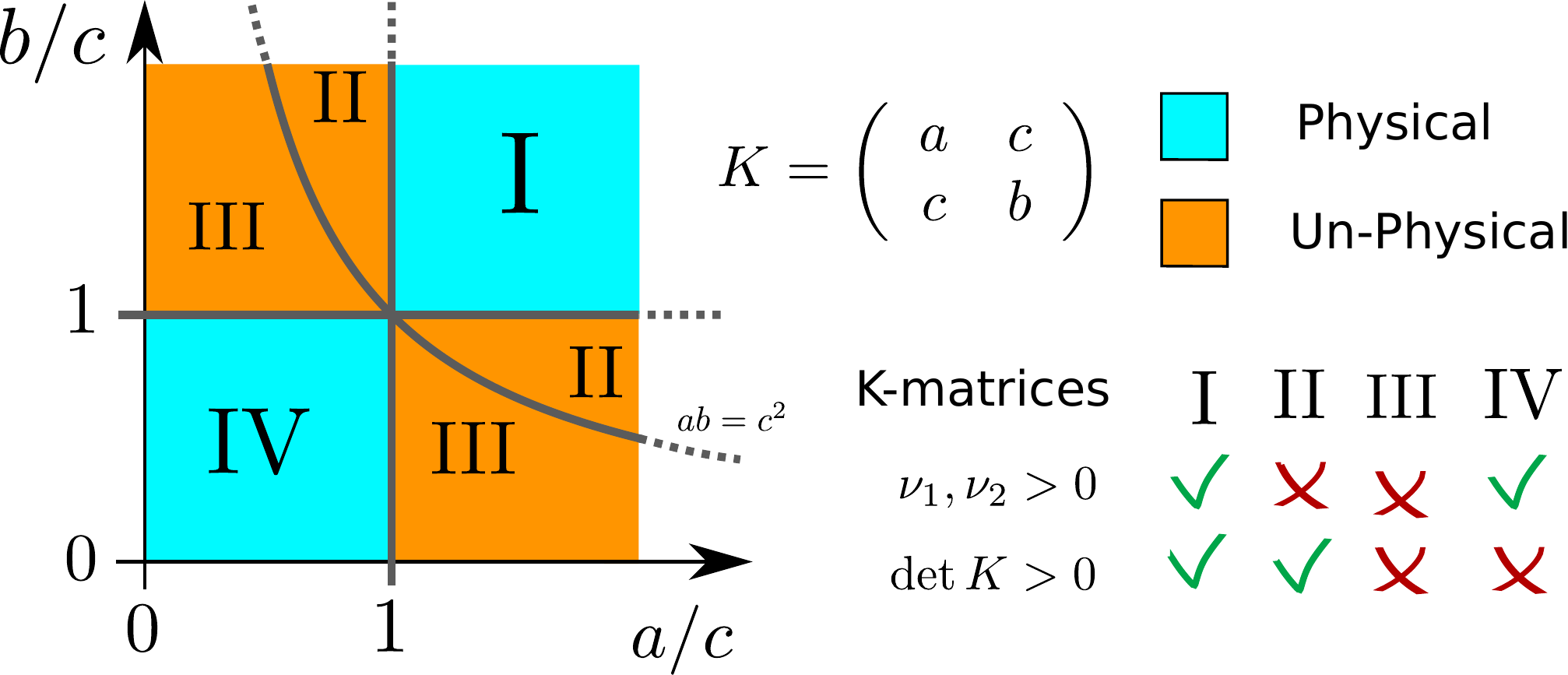}
  \caption{The classification of the space of possible K-matrices of the form abc, \ie $K=\left(\begin{array}{cc} a & c\\ c & b \end{array}\right)$, with $a,b,c>0$.
    Region I contains the positive definite K-matrices, which can also be modelled with CFT techniques. Regions II has two positive eigenvalues, but one of the filling fractions is negative. Region three has one negative eigenvalue and one negative filling fraction, $\nu_j$. Both regions II and III are unphysical. Region IV has one negative eigenvalue, but both filling fractions are positive and can, therefore, be physical.
    The CFT approach gives holomorphic CoM pieces in regions I and II whereas the monodromy matching only does the same in region I.
    On the other hand, the CFT cannot produce a CoM piece in region IV which the monodromy matching can.}
  \label{Fig:K-matrix-regions}
\end{figure}

\subsection{A physical requirement for K-matrices}

Before moving on to the explicit realisations of the selected states, let us first discuss which K-matrices are physically realisable.
Taking equation \eqref{eq:CFTCoM} at face-value, one would conclude that a CFT construction is possible as long as all the eigenvalues of the K-matrix are positive. Therefore, we should also be able to use the monodromy matching to find the corresponding CoM terms.
However, considering, \eg the K-matrix $K=\left(\begin{array}{cc} 1 & 3\\ 3 & 11 \end{array}\right)$ we find an apparent contradiction.
In this case, the eigenvalues, $\lambda=6\pm\sqrt{34}$, are both positive, yielding a well-defined (holomorphic) CoM function from the CFT construction.
At the same time, however, the monodromy matching would predict the structure
\begin{equation}
    \label{CoMMCMonodromy1311}
    \mathcal{G}(\vec{Z}|\tau) = \frac{\vartheta^3(Z_1+Z_2|\tau)\vartheta^8(Z_2|\tau)}{\vartheta^2(Z_1|\tau)}~,
\end{equation}
which has poles for the $Z_1$ variable.
The solution to this apparent paradox is that this K-matrix, in a specific sense, does not correspond to a physical state.
Although the eigenvalues are both positive, we find that filling fractions are 4 and -1, respectively, which predicts a negative amount of particles of the second species.
Since a negative amount of particles is an unphysical configuration, it explains the mismatch between the two realizations (CFT and monodromy matching).
In Figure \ref{Fig:K-matrix-regions}, we show how the sign of the determinant is related to the sign of the filling fractions and which K-matrices allow for positive densities for all its species.

\section{Explicit realizations of wave functions on the torus}
\label{sec:V-Realizations}

In this section we will use \eqref{MMAnsatzGeneral} and provide CoM pieces for a selection of K-matrices.
We will first discuss the Laughlin state, then move to the positive definite case of 331 state (or More-Read state when anti-symmetrized). Finally we will relax the requirement of positive-definiteness, and focus on the 113 states.
In the first case we can check our results against the ones obtained via correlators of conformal field theories (CFTs)~\cite{hsbhk08,hhrv09,hhv09}, but in the latter no CFTs exist.
We mention in passing that if one relaxes the condition to have a fully holomorphic Jastrow factor $J$ then the CFT correspondence is able to handle non-positive K-matrices. This however, comes with the cost of creating states with large components in higher Landau levels\cite{svh11_2}, \ie very far from the LLL. For these states projection is also prohibitively expensive.

\subsection{Laughlin states}
\label{Laughlin states}
Laughlin states are known to describe fQHE states at the LLL for filling fractions $1/q$, with $q$ odd, as a condensate of spin-polarized electrons with the lowest relative orbital angular momentum being $q$~\cite{l83}.

Hence, on the torus, the Jastrow part reads
\begin{equation}
    \label{JFL1}
    J(z_1,\dots,z_N|\tau) = \prod_{i<j=1}^N\left\{\elliptic{1/2}{1/2}{z_i-z_j}\tau\right\}^q.
\end{equation}
This formulations already implements the correct fermionic exclusion statistics and the short-range behaviour, which must be equivalent to the planar formulation of the problem~\cite{hr85}.
As a consequence (see Appendix \ref{App: Laughlin state} for details), the CoM piece (for odd $N_e$) is constrained to have the following form
\begin{equation}
    \label{CoML1}
    \mathcal{F}_s(Z|\tau)=\frac{1}{\eta(\tau)}\elliptic{s/q}0{qZ}{q\tau},
\end{equation}
where we can directly read of the $q$-fold degeneracy of the ground state through the label $s$.

\begin{figure}
    \centering
    \includegraphics[width=0.9\columnwidth]{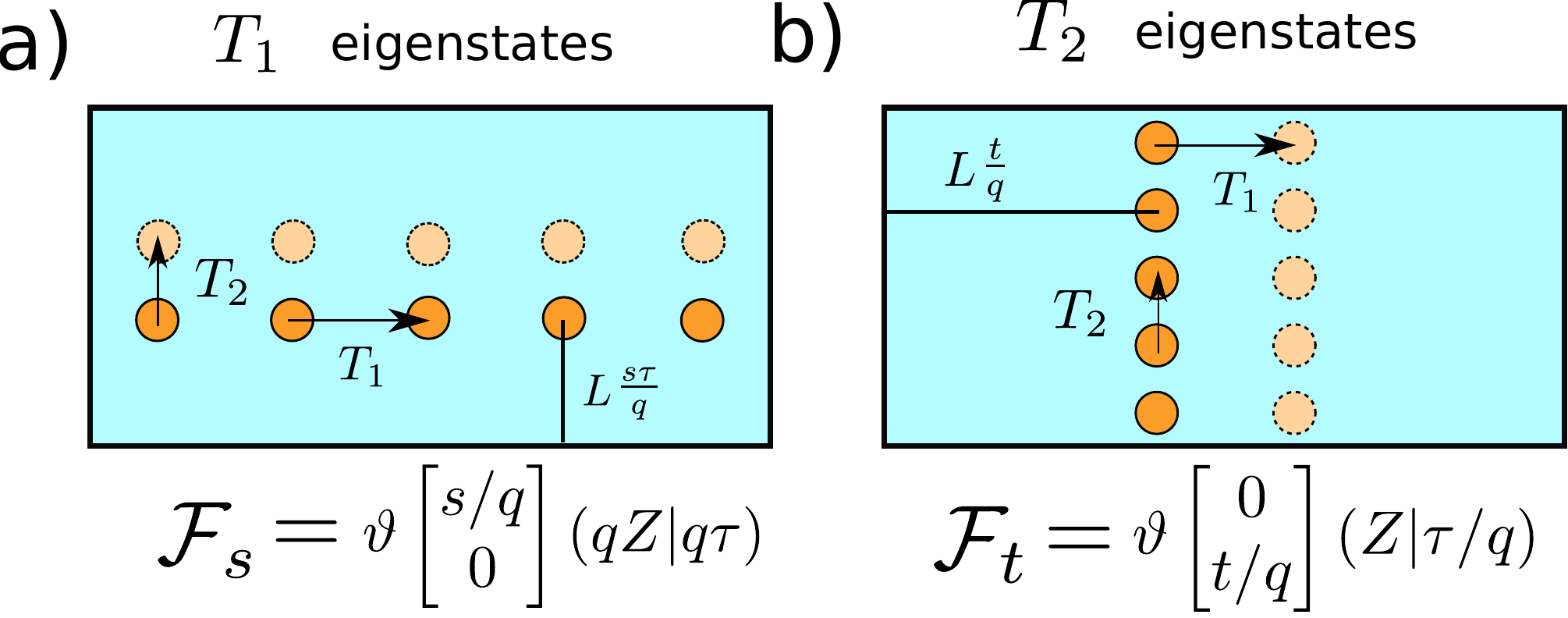}
    \caption{The structure of the zeros of the Laughlin CoM term, $\mathcal F$, for a) eigenstates of $T_1$ and b) eigenstates of $T_2$. In both cases the zeros of $\mathcal F$ sit on a one-dimensional line, in the direction indicated by the translation of $T_j$. The action of $T_j$ move each zero onto the next one, mapping the state onto itself.
      The other operator $T_{\neq j}$ will move the row of zeros in the orthogonal direction, leading to a new state (with a different quantum number).
    }
    \label{Fig:ZeroLaughlin}
\end{figure}

The Laughlin states CoM in \eqref{CoML1} can also be cast in the form of \eqref{MMAnsatzGeneral}, if we use the structure of the zeros of $\mathcal{F}_s(Z|\tau)$.
We know that $\mathcal{F}_s(Z|\tau)$ is an eigenstate of the $T_1$ operator with precisely $q$ zeros, and the action of $T_1$ is to send $Z\to Z+1/q$.
The zeros of $\mathcal{F}_s(Z|\tau)$ must therefore lie on a line parallel with the real axis and with a separation of $1/q$.
This can only happen if the $j$:th zero of $\mathcal{F}_s(Z|\tau)$ is at position $Z_j = \frac{j}{q} + \frac{s}{q}\tau$ for $j=0,\dots,q-1$.
Hence we see that $\mathcal{F}_s(Z|\tau)$ can be written as the product
\begin{equation}
    \label{CoMLMMA}
    \mathcal{F}_s(Z|\tau) \propto \prod_{j=0}^{q-1} \elliptic{s/q}{j/q}Z\tau~,
\end{equation}
where the proportionality is there at account for non-trivial prefactors which depend on $\tau$.
The structure of the zeros is illustrated in Figure \ref{Fig:ZeroLaughlin} for both eigenstates of the a) $T_1$-, and b) $T_2$-operators.

Laughlin states can be generalized to describe different filling fractions $\nu$ in two different ways: either via composite fermions~\cite{jk97,jk97_2}, or via multi-component (or hierarchical) states~\cite{h83}.
In the next sections we analyze the latter realization of fQHE on the torus.

\subsection{Multi-component states with positive definite K-matrix}
\label{Multicomponent states with positive definite K-matrix}

So far, we have discussed states consisting of a single component, \ie all the particles in the trial wave functions are on the same footing.
By relaxing this condition, we can construct multi-component wave functions that rely on the following assumption: we can define a ``layered" iQHE ground state, \ie a state made up of $n$ different particle species where the repulsion among the same kind of particles is different from the repulsion between particles of different species.

We can write down an effective Jastrow part also in the multi-component case as the CFT case \eqref{JastrowCFT}, where after anti-symmetrization $\A$ in the full wave function \eqref{QHEwavefunction} all the particles are indistinguishable.
This is desirable when the proposed wave function aims to describe a single fQHE liquid.
This Jastrow part straightforwardly generalizes the construction for the Laughlin state \eqref{JFL1} for multiple species and allows for different values of the filling fraction $\nu$.
When the K-matrix is diagonal, equation~\eqref{JastrowCFT} reduces to a product of different Laughlin states' Jastrow parts since we are then just considering a collection of decoupled Laughlin states.
On the other hand, one can still consider the states without anti-symmetrization as realizations of multi-layered iQHE systems, which are known in the literature as Halperin states~\cite{h83_3}.
They can represent a system composed of different fQHE states that exhibit repulsion among their fundamental components, as in different layers of fQHE with coupling among them~\cite{qjm89}.
These different species usually share the same orbital, and the interactions do not distinguish among different orbital states.
However, it has been shown in bilayer graphene that other fQHE states can emerge from pairing different orbital states from different layers~\cite{hbfgwtjz25}.

One can distinguish two qualitatively different cases, \ie when the K-matrix's diagonal terms are larger or smaller than off-diagonal ones.
In the first case, a particle will experience stronger repulsion from particles of the same species instead of the one done by those from other species.
As a consequence, particles will distribute uniformly on the torus, helped by anti-symmetrization.
In the second one, a particle is more likely to be close to particles from the same species rather than those from different species.
Namely, particles will tend to cluster and separate from other species before anti-symmetrization.
It was argued in~\cite{sb24} that, after anti-symmetrization, the clustering remains present.

When the K-matrix is positive definite, the CoM piece can be expressed as a holomorphic function given by Eq. (\ref{eq:CFTwfn-Kmatrix}) earlier in the text. 
From it, it is straightforward to prove that there are exactly $|\det(K)|$ different ground states.
As a side remark, the formulation in  \eqref{eq:CFTwfn-Kmatrix} is an eigenstate of $T_1$, but not necessarily of $T_2^q$.
To achieve that as well, linear combinations of CoM pieces with appropriate phase factors may need to be constructed.

The K-matrix formulation has the advantage of being directly derived from knowing the Jastrow piece \eqref{JastrowCFT}; however, an equivalent result can be derived in terms of one-dimensional theta functions through monodromy matching.
An instance of that is the state
\begin{equation}
    K = \begin{pmatrix}
        3 & 1 \\
        1 & 3
    \end{pmatrix} ,
\end{equation}
which was also used as an example in Section \ref{sec:Monodromy_matching}. The K-matrix represents the Moore-Read Pfaffian state at filling $\nu=5/2$~\cite{mr91,abgs98}.
The state is known to be a good fQHE ground state; its excitations are non-Abelian anyons, and it is a candidate to describe the ground state of the fQHE at half-filling.
In this case, particles do not cluster, and an energetic gap is present, \ie the state is incompressible.

Imposing the correct boundary conditions leads to a CoM factor of the form 
\begin{eqnarray}
    \label{CoMMCMonodromy331}
    \G_{331}(\vec{Z}|\tau) &=& \left\{\elliptic{\alpha_1}{\beta_1}{Z_1}\tau\elliptic{\alpha_2}{\beta_2}{Z_2}\tau\right\}^2
    \nonumber\\
    &&\times\elliptic{\alpha_1+\alpha_2}{\beta_1+\beta_2}{Z_1+Z_2}\tau,
\end{eqnarray}
and the full expression can be found in the Appendix \ref{App: 331 state}.
We have numerically verified that the expression above can always be decomposed into only eight distinct basis states despite having numerous free parameters. The eight states are precisely the ones obtained in terms of multi-dimensional theta functions using the CFT approach. 
See Appendix \ref{App:CoM-basis} for more details.

\subsection{Multi-component states with non-positive definite K-matrix}
\label{Multicomponent states with non-positive definite K-matrix}
A simple instance of multi-component states with non-positive definite K-matrix can be considered the 113 state.
This case is described by a very similar K-matrix to the 331 case
\begin{equation}
    K = \begin{pmatrix}
        1 & 3 \\
        3 & 1
    \end{pmatrix} ,
\end{equation}
in which the off-diagonal elements are exchanged with the diagonal ones.
The Halperin construction is known to exhibit a breakdown of the plasma analogy~\cite{qjm89}, as long as it represent a double layer of iQHE devices in which the particles contained in one layer repel more the ones contained in the other, compared to how the constituents on the single layer repel themselves, inducing clustering in the system.
The anti-symmetrized version of such state on the other hand would represent the so called $A$-phase in~\cite{ddm23} with analogous filling fraction $\nu=1/2$ to the previous positive-definite case.
In spite of this similarities with the previous cases, it describes completely different physics, since now particles from different species are repelling more than constituents from the same kind, but anti-symmetrization might eliminate the difference between the two particle species.

Moreover, in this case, the CoM piece cannot be written as any linear combination of multi-dimensional theta functions using \eqref{eq:CFTwfn-Kmatrix} since the eigenvalues of the K-matrix are not positive, and therefore the series representation is not convergent (see Appendix \ref{App: Series representation} for details).
The non-convergence is evident in the solution to \eqref{eq:Monodromy_constraints} as well: in this case $d_{1,2}=d_{2,1}=3$ and $d_1=d_2=-2$ and the negative values of $d_1$ and $d_2$ will have dramatic effects on the CoM piece
\begin{equation}
    \label{CoMMCMonodromy113}
    \G(\vec{Z}|\tau) = \frac{\vartheta^3(Z_1+Z_2|\tau)}{\vartheta^2(Z_1|\tau)\vartheta^2(Z_2|\tau)}~,
\end{equation}
by introducing isolated singularities in the wave function (for simplicity we dropped the dependence on the parameters $\alpha$ and $\beta$ in this expression).

A way to regularize such singularities is to map the problem into its non-holomorphic description, as shown in Section \ref{Non-positive definite K-matrices}. 
In this case, the 113 state turns the modified CoM term into
\begin{equation}
    \label{CoMMCAH113}
    \G_{113}(\vec{Z}) = G_{AH}(\vec{Y})\vartheta^3(Z_1+Z_2)\left[\vartheta^2(Z_1)\vartheta^2(Z_2)\right]^{*},
\end{equation}
with the extra Gaussian factor
\begin{equation}
    \label{GaussCorr113}
    G_{AH}(\vec{Y}) = e^{-2\pi\tau_2(2Y_1^2+2Y_2^2)}~.
\end{equation}
In this state, the Jastrow factor is unchanged as compared with the state in \eqref{CoMMCMonodromy113}, although the CoM is now pierced by opposite magnetic fluxes that stabilize the configuration.
Under another, complementary, point of view, we have traded a local non-analyticity of the wave function for a global one.

One may wonder if the use of non-holomorphic wave functions will result in a well defined Hall viscosity.
Hall viscosity from non-LLL states has previously been computed in Ref.~\cite{pfj20} for composite fermions, showing agreement with results coming from LLL state for different fillings.
Hence we expect these non-holomorphic wave functions to provide still a finite result for fQHE states.
We are now in the position to investigate the $A$-phase by computing its Hall viscosity.


\section{Hall viscosity and numerical analysis}
\label{sec:Hall viscosity}
In this chapter, we will numerically test the proposed wave functions created in previous chapters by computing the Hall viscosity, $\eta^H$, of the states.
The Hall viscosity is a manifestation of the underlying topological order of a system, in that it can be linked to the average orbital spin, $\sbar$, which in turn is related to the shift, $\mathcal S$, of the system when put on a sphere.

The description of fQHE on a torus allows for a straightforward definition of the Hall viscosity by means of computing the adiabatic curvature on the ground state manifold parameterised by the modular parameter $\tau$.
In our case, since we have real-space wave functions, we use Monte Carlo sampling, given that the computation of the scalar product between different wave functions is required.
Through out this section we are considering wave functions that are eigenstates of the two CoM translation operators \eqref{eq:T-CoM}, $T_1$ and $T_2^q$\footnote{Note that since the ground state manifold in the torus is degenerate, in principle one needs to compute the berry curvature matrix, however, by using the eigenstates of $T_1$ and $T_2^q$ we separate the different momentum sectors and focus on one without loss of generality}.

\subsection{Hall viscosity}
\label{subsec:Hall viscosity}
Since the magnetic field in integer and fractional QHE breaks time reversal of the system, it is expected that the stress tensor contains contributions also from the anti-symmetric part of the viscous response $\eta^H$~\cite{asz95}.
In contrast to the symmetric part, such terms are not related to dissipation in any form; in particular, when the ground state of a gapped system at zero temperature is probed, any dissipation is absent.
It was shown by Read~\cite{r09} that for states constructed via conformal blocks, $\eta^H$ is proportional to the average orbital spin, which is in turn proportional to the shift on the sphere and makes it a topological invariant for clean systems.
In general, the same argument can be extended to other FQHE states as long as the Laughlin plasma analogy holds~\cite{r09}.

We follow the original construction and relate the Hall viscosity to the Berry curvature $\mathcal{F}_{\tau\bar{\tau}} = \ii\partial_{\bar{\tau}}A_{\tau} - \ii\partial_{\tau}A_{\bar{\tau}}$ as
\begin{equation}
    \label{eq: HallViscosity}
    \eta^H = -\frac{2\tau_2^2}{A}\mathcal{F}_{\tau\bar{\tau}} ,
\end{equation}
where we restored the area of the torus $A=\tau_2 L^2$ for completeness.
We recall that the Berry potential is related to the modular parameter $\tau$ (and its complex conjugate) as
\begin{equation}
    \label{eq: BerryPot}
    A_{\tau} = \ii\langle\Psi(\tau)|\partial_\tau\Psi(\tau)\rangle.
\end{equation}

It is customary to relate Hall viscosity with average orbital spin $\bar{s}$ and electronic density $\bar{n}$
\begin{equation}
    \label{OrbitSpin}
    \eta^H = \frac{1}{2}\bar{n}\bar{s}\hbar.
\end{equation}

In order to analytically compute the Hall viscosity, $\eta^H$, it is vital to not only know the wave function; one also needs control of the normalization as a function of $\tau$ (and $\bar\tau$). However, in most cases, as is here, the norm is not known, but one can reach approximate result by invoicing further assumptions, such as the plasma analogy.

\begin{figure}
  \centering
  \includegraphics[width=0.45\columnwidth]{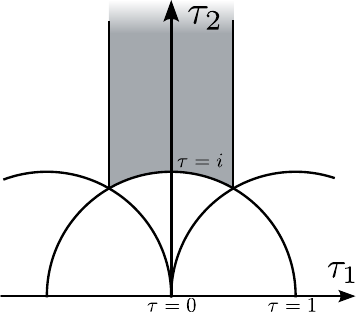}
  \includegraphics[width=0.45\columnwidth]{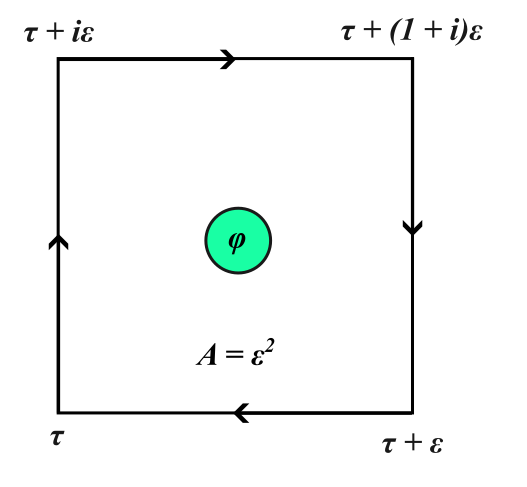}
  \caption{(left) The modular plane, parameterized by $\tau$. In grey, the fundamental domain bounded by $|\tau|>1$ and $|\tau_1|<\frac12$. Special points in the place are marked as a guide to the eye. (right) Berry flux $\varphi$ on a plaquette of area $\varepsilon^2$ with a vertex placed in $\tau$. The arrows indicate the direction of the circuitation on the path that encloses the square area.}
  \label{Fig:BerryFlux}
\end{figure}

Finally, we briefly comment on the geometric meaning of the previous quantity.
First, we remark that the absence of a well-defined Hall viscosity signals the presence of dissipation in the system: this might be, for instance, caused by a compressibility of the bulk in the wave function formulation (\ie a vanishing gap).
However, even for a well-defined QHE state, the viscosity is not perfectly quantized, since the parameter space is not compact.
As a matter of fact, the manifold is the ground state bundle on the fundamental domain of the modular group, whose representations are always unbounded. For instance, on the complex plane, the fundamental domain extends towards infinite values along the imaginary direction; see Fig. \ref{Fig:BerryFlux}, left panel.

The Hall viscosity can be intuitively related to the action of a deformation of the underlying torus geometry. Still, this interpretation lacks rigour: more precisely, it is the linear response to the strain induced by a generalized force that couples to the adiabatic curvature, and thus it does not imply a change in the metric of the physical space.

\begin{figure}
  \centering
  \includegraphics[width=\columnwidth]{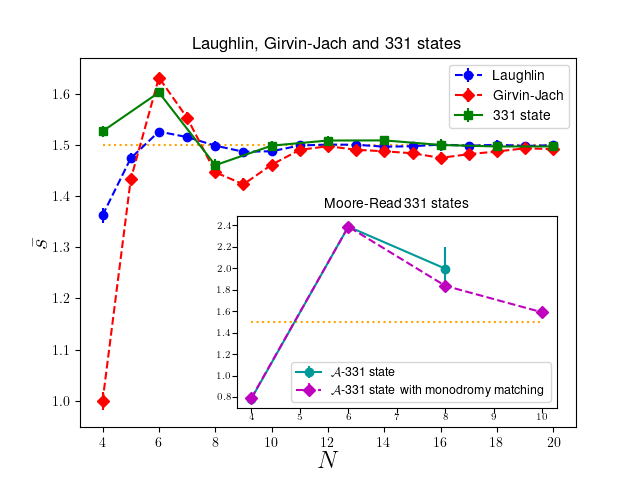}
  \caption{Hall viscosity encoded by average orbital spin, $\sbar$ of the Laughlin state $q=3$, it's Girvin-Jach excited state with $k=1$, and the un-symmetrized 331 state.
    Dashed orange line show the expected value of $\bar{s} = 3/2$ computed using CFT arguments~\cite{r09}.\\
    {\em inset:} $\sbar$ for the $\A$-331 state using CFT methods and monodromy matching. As expected the two approaches give identical results (up to numerical uncertainty).
}
  \label{Fig:HallViscosityL-GJ}
\end{figure}

\subsection{Monte Carlo sampling}
\label{Monte Carlo sampling}
We compute the viscosity at finite $N$ by estimating the Berry flux through a small square of side $\varepsilon$, with a vertex placed at $\tau$; and extract $\mathcal{F}_{\tau\bar{\tau}}$ by discretizing the derivatives in \eqref{eq: HallViscosity}.
The Berry curvature then reads
\begin{multline}
    \label{BerryCurvNum}
    \mathcal{F}_{\tau\bar{\tau}}\varepsilon^2\approx
    \langle\Psi(\tau)|\Psi(\tau+\varepsilon)\rangle 
    + \langle\Psi(\tau+\varepsilon)|\Psi(\tau+\varepsilon+\ii\varepsilon)\rangle \\
    - \langle\Psi(\tau+\ii\varepsilon)|\Psi(\tau+\varepsilon+\ii\varepsilon)\rangle - \langle\Psi(\tau)|\Psi(\tau+\ii\varepsilon)\rangle,
\end{multline}
where moved $\varepsilon^2$ to the right-hand side for ascetics.
Geometrically, the right-hand side of the expression represents the circulation of the Berry connection $\A_\tau$ on the infinitesimal square; see Figure~\ref{Fig:BerryFlux} (right side) for an illustration.
To evaluate this expression, we sample the wave function at $\tau$ using the Monte Carlo algorithm, and then we generate the related wave functions at the points $\tau + \varepsilon$, $\tau + \ii\varepsilon$ and $\tau + \varepsilon + \ii\varepsilon$ from the coordinates of the first sampling.
The overlap is taken between states with torus characteristics corresponding to the value attached to adjacent vertices.
We estimate the errors induced by Monte Carlo sampling by repeating the Hall viscosity calculation across multiple MC datasets.

\subsection{Numerical results}
\label{Numerical results}

We here report on the numerical results of the proposed wave functions.
For brevity, we will refer to the anti-symmetrized states of 331 and 113 as $\A$-331 and $\A$-113, respectively.

\subsubsection{Hall Viscosity of regular states}
To begin with, we fix the geometry at $\tau=\i$, and compute the Hall viscosity for a selection of regular wave functions as a function of the particle number. In Figure~\ref{Fig:HallViscosityL-GJ}, we begin with the $\nu=1/3$ Laughlin state, its counterpart generated by the Girvin-Jach construction for $k=1$ (defined in eq.~\eqref{EQ:GJ-WF}), and the multi-component 331-Halperin state.
We expect these three states to share the same value of the orbital spin $\bar{s}$, which can be derived from plasma analogy.
Moreover, we do not need to perform anti-symmetrization numerically in any of these states; the Laughlin and Girvin-Jach states are already antisymmetric by construction, and the Halperin state describes layered materials where electrons from different subsystems are distinguishable.
This fact allows us to use Monte Carlo sampling for significantly larger sizes $N$ than with explicit antisymmetrization, and to obtain better convergence of our results.

We observe sizable fluctuations in $ \bar{s}$ for small $ N$, especially for the Girvin-Jach state.
Nonetheless, both curves quickly approach the analytically expected value of $\simeq 1.5$ for $N>10$~\cite{rr08}.
This result serves as both a sanity check for our code and a confirmation of the statement that Hall viscosity can also be extracted from higher Landau levels, as seen in the unprojected state of Laughlin's Girvin-Jach version with $k=1$.
This second construction, when projected to the LLL, is closer to the actual Coulomb ground state~\cite{ffms16} but shows weaker stability in its result.
Also, the multi-component 331-Halperin state converges to the expected viscosity as the number of particles increases.
From the convergence of the three, we conclude that already for $10$ particles, Hall viscosity can be inferred in these cases.

Secondly, we introduced anti-symmetrization to obtain the $\A$-331 state (or More-Read state); see the inset of Figure~\ref{Fig:HallViscosityL-GJ}.
In the figure, we compare $\sbar$ for $\A$-331, constructed using the CFT method and the monodromy matching.
As anti-symmetrization is an expensive operation, we did not push the numerics too far and stopped at only 10 particles.
We observe 1) that the results are the same for the two approaches, which we also expect, given that the two methods give identical formulations of the same wave function.
2) Due to the smaller system sizes, the value of $\sbar$ has not entirely settled at the expected $\sbar=3/2$, but the tendency is clear that it will.

The fact that both the 331 and $\A$-331 states share the same Hall viscosity suggests that anti-symmetrization does not alter the physical response of the stress tensor. Using the plasma analogy~\cite{r09}, this can also be derived explicitly.

\subsubsection{Hall Viscosity of 113 and $\A$-113}
We now turn our attention to the non-positive definite cases of $K=(113)$, with and without anti-symmetrisation.
It is a well-known fact from the sphere that the 113 state shows self-separation and does not constitute a good FQH wave function~\cite{sb24}.
Therefore, it is instructive to examine the viscosity signature of this behaviour on the torus.

\begin{figure}[t]
  \includegraphics[width=0.5\textwidth]{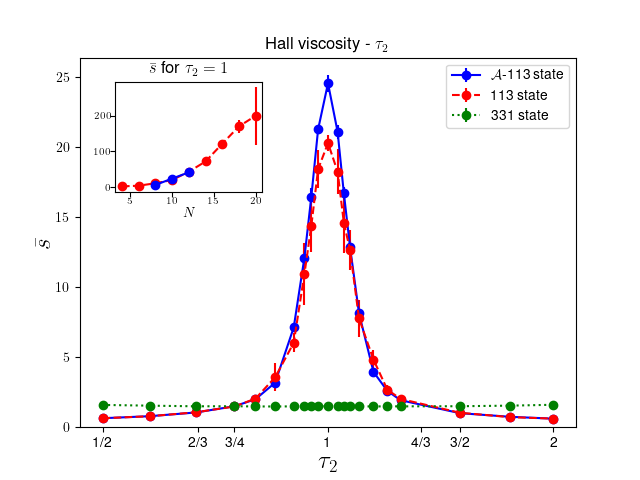}
  \caption {Average orbital spin $\sbar$ for the states 113, $\A$-113 and 331, with $N=10$ and $\tau=\i \tau_2$.
    Both the 113 and $\A$-113 states show a prominent peak in $\sbar$ around $\tau=\i$, whereas 331 does not.
    The behaviour signals a strong dependence on the geometry parameter $\tau$, which should not be the case for a well-developed topological phase.
    A logarithmic $\tau_2$ axis has been chosen to highlight the modular independence of the results under the transformation $\tau\to-1/\tau$ (or here $\tau_2\to1/\tau_2$). \\
    {\em inset:} $\sbar$ for the states 113 and $\A$-113 with $\tau=\i$ fixed and varying $N$. As $N$ increases, so does $\sbar$, showing that there is likely no thermodynamically stable value for $\sbar$ in these systems.
    }
  \label{Fig:HallViscosityTau}
\end{figure}

Here, we keep the particle number $N=10$ fixed and vary the imaginary part of the geometry parameter $\tau$.
The result can be seen in Figure~\ref{Fig:HallViscosityTau}, where we plot $\sbar$ for 113, $\A$-113 and the 331 state.
We find that for 113 and $\A$-113, changing $\tau_2$ creates large fluctuations in $\sbar$ with a pronounced peak at $\tau=\i$. The same peak is not present for 331, which stays approximately constant at $\sbar\approx 3/2$.
Furthermore, as seen in the figure inset, the peak at $\tau=\i$ increases with $N$, suggesting that there is no thermodynamically stable value of $\sbar$ in these systems.

The above-described behaviour is in stark contrast to good fQHE states, where the viscous response is independent from the geometric details of the configuration (\ie, the value of the torus parameter $\tau$).
The nature of the response is intimately related to the screening in the plasma analogy, or analogously to the presence of a gap in the neutral sector~\cite{rr08}.
The observation of a $\tau$-dependence is consistent with the idea of a phase-separated liquid, since the system cannot support screening. Further, gapless excitations can occur along the boundary separating the two phases (see Appendix~\ref{App: Boundary energy of self-separated plasma} for a discussion).
As such, the system would be sensitive to the shape of the geometry it is in, and the viscosity would reflect that.

\begin{figure}[t]
  \centering
  \includegraphics[width=0.45\textwidth]{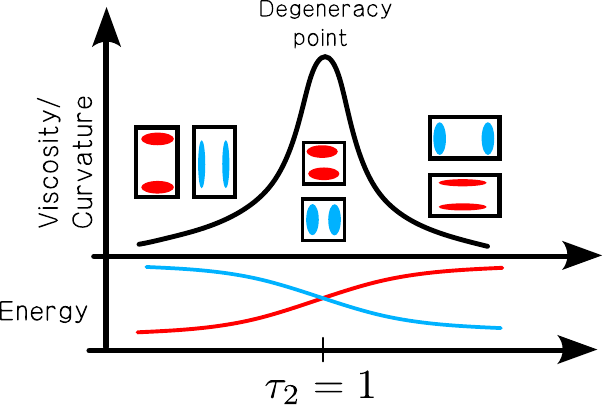}
  \caption{Qualitative sketch of the mechanism leading to a peak in viscosity as $\tau=\i$.
    The top plot shows the system's viscosity as a function of $\tau_2$; the bottom shows the energies of two possible configurations of self-separated liquids.
    The insets illustrate how the two possible states (in red and blue) could be arranged on the torus.
    The state with the largest separation between liquids will have the smallest energy; still, at $\tau_2=1$, the two configurations will have the same energy.
    At the degeneracy point, the system will transition from the red to the blue state, resulting in a macroscopic change in the ground state. That change will, in turn, be picked up by the viscosity as a divergence.}
  \label{Fig:Visc_Divergence}
\end{figure}

We present here an intuitive picture to explain why a self-separated liquid would exhibit a divergence at the square-torus point ($\tau=\i$).
As sketched in Figure~\ref{Fig:Visc_Divergence}, we can consider the torus in the regimes with $\tau_2<1$, where the height of the rectangle is larger than its width, and the opposite case with $\tau_2>1$.
On both sides, the more stable configuration is the one in which the two liquids are as far apart as possible, placing the two electronic puddles along the shorter sides.
The repulsion means that, in the first case, the ground state is the state whose phase-separation boundary splits the rectangle along its width. In the second case, it splits the rectangle along its height instead.
It is then straightforward to see that at the squared-torus point $\tau_2=1$, the two configurations are energetically equivalent, causing level crossing in the system.
We now note that the Hall viscosity is connected with the rate at which the wave-function changes with respect to $\tau$,
via the Berry potential~\eqref{eq: BerryPot}.
As such, a level crossing will cause a sharp transition from one state to another, which will be reflected as a divergence in the viscosity.
Taking the finite differences instead of the exact derivatives regularises this infinity, but we can identify it by studying how $\sbar$ depends on the parameter regularisation parameter $\epsilon$ (see Appendix~\ref{App: The divergence} for more details).

We conclude that neither the 113 nor $\A$-113 states provide a well-defined value of the Hall viscosity, and therefore cannot represent fractional quantum Hall liquids.
In the following sub-section, we will give further corroborating evidence that both the 113 and $\A$-113 states are self-separating, also on the torus.

\subsubsection{Clustering diagnostics}
We will now investigate the $G$-distribution of the particles using the techniques introduced in Ref.~\cite{sb24} for the sphere, here modified for the torus.
We will focus on the comparison of the 331, 113, and $\A$-113 states.

The $G$-statistic is a measure of the internal distribution of distances between particles in a fluid. For a more extensive discussion, we refer to the original paper; here, we only repeat the main features.
Also, some technical differences are listed in Appendix~\ref{App: Geometric artifacts in cluster diagnostics}.

First, let $R(k)$ be the average distance from any particle to the $k$:th closest particle to that one.
It is true that, for $N$ equally spaced particles
\begin{equation}
    [R(k)]^2 \approx [R_0(k)]^2 =  \frac{k}{\pi \rho}~,
\end{equation}
where $\rho$ is the density of particles.
We now construct
\[G(k) = [R(k+1)]^2 - [R(k)]^2 \]
as the squared difference between then $k+1$:th and the $k$:th average distances.
For a system where the particles are uncorrelated, $G$ will be roughly constant with $G\approx 1/(\pi\rho)$.
If there is a peak in $G$ around some $k$, it means there is an anomalous increase in the distance to the $(k+1)$:th closest particle.

In this work, instead of considering $G(k)$, we will focus on $R(k)$ itself. The difference is that computing G(k) in the torus can lead to geometric artifacts, which impede interpretation; see Appendix~\ref{App: Geometric artifacts in cluster diagnostics} for a further discussion.
We will report the ratio between the average distance $R(k)$ to particle $k$, and the value expected from a homogeneous spatial distribution on a torus $R_0(k)$.

\begin{figure}[t]
    \centering\includegraphics[width=\columnwidth]{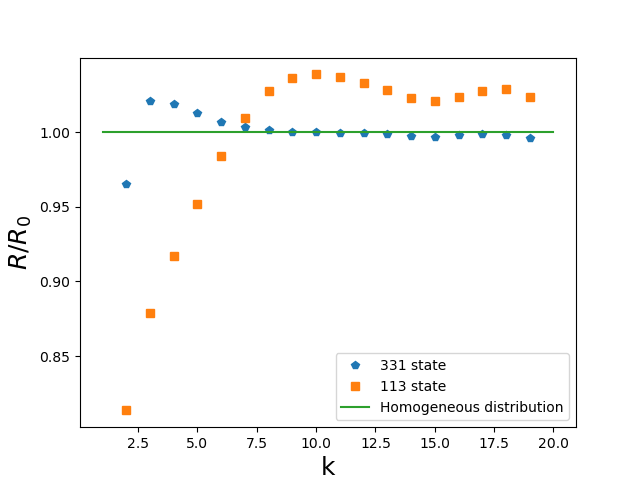}
    \caption{
      Ratio between the average distance to the $k$:th particle $R(k)$ and the value expected from homogeneous spatial distribution on a torus $R_0(k)$ for the 331 and 113 Halperin states (blue and orange dots respectively) and $N=20$. The green solid line represents the distribution of perfect homogeneous fluids, which is the case for the 331 state at large $k$. The 113 state shows a different behavior in which particles close to the reference particle are closer among themselves than in a homogeneous liquid, and particles far away are farther away than in the homogeneous case.
    }
    \label{Fig:ClusterHal}
\end{figure}

We begin by comparing the $R$ clustering statistics for the 113 and 331 states at $N=20$ on the square torus; see Figure~\ref{Fig:ClusterHal}.
We see that the 331 and 113 states show very different behaviors in the $R$-statistic.
While the 331 state (blue) is almost constant, indicating a homogeneous fluid, the 113 state shows its closest particles much closer than one would naively expect. Likewise, the particles further away are farther away than one would expect for a homogeneous fluid.
This behaviour is to be expected, given that the 113 Halperin state is self-separating on the sphere.
The underlying mechanism is that particles from different groups repel more than those of the same kind, leading to the components of one layer always being far apart from those in the other layer.

\begin{figure}[t]
    \centering
    \includegraphics[width=\columnwidth]{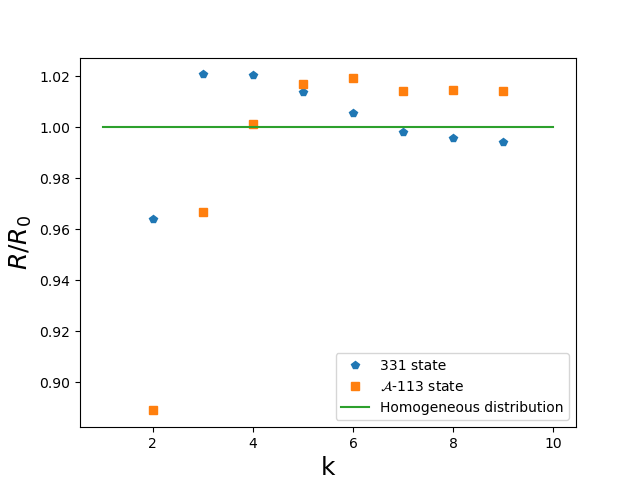}
    \caption{
    Same setup as in Figure \ref{Fig:ClusterHal}, but now for the 331 Halperin state (blue dots) and the anti-symmetrized 113  state (orange dots) with $N=10$.
    Also here, even after anti-symmetrization, the $\A$-113 state does show a self-separation of the particles in the fluid. The 113 state is less clean than for $N=20$, showing clear finite-size effects for $k<6$.
}
    \label{Fig:ClusterPH}
\end{figure}

Let us now perform the same computation for the anti-symmetrized state $\A$-113, compared with the Halperin state 331, but now at $N=10$; see Figure \ref{Fig:ClusterPH}.
In this case, the argument of separated groups partially breaks down because particles are now indistinguishable.
However, as in Figure~\ref {Fig:ClusterHal}, we also observe the same qualitative behavior, albeit less pronounced due to the smaller system size.
It therefore appears that anti-symmetrization alone is not sufficient to prevent clustering of the ground state.
This result is consistent with the hypothesis advanced in the study of the Hall viscosity. It allows identification of particle clustering as the mechanism that destroys both screening in the system and the topological properties of the state.

\begin{figure*}
    \centering
    \includegraphics[width=\linewidth]{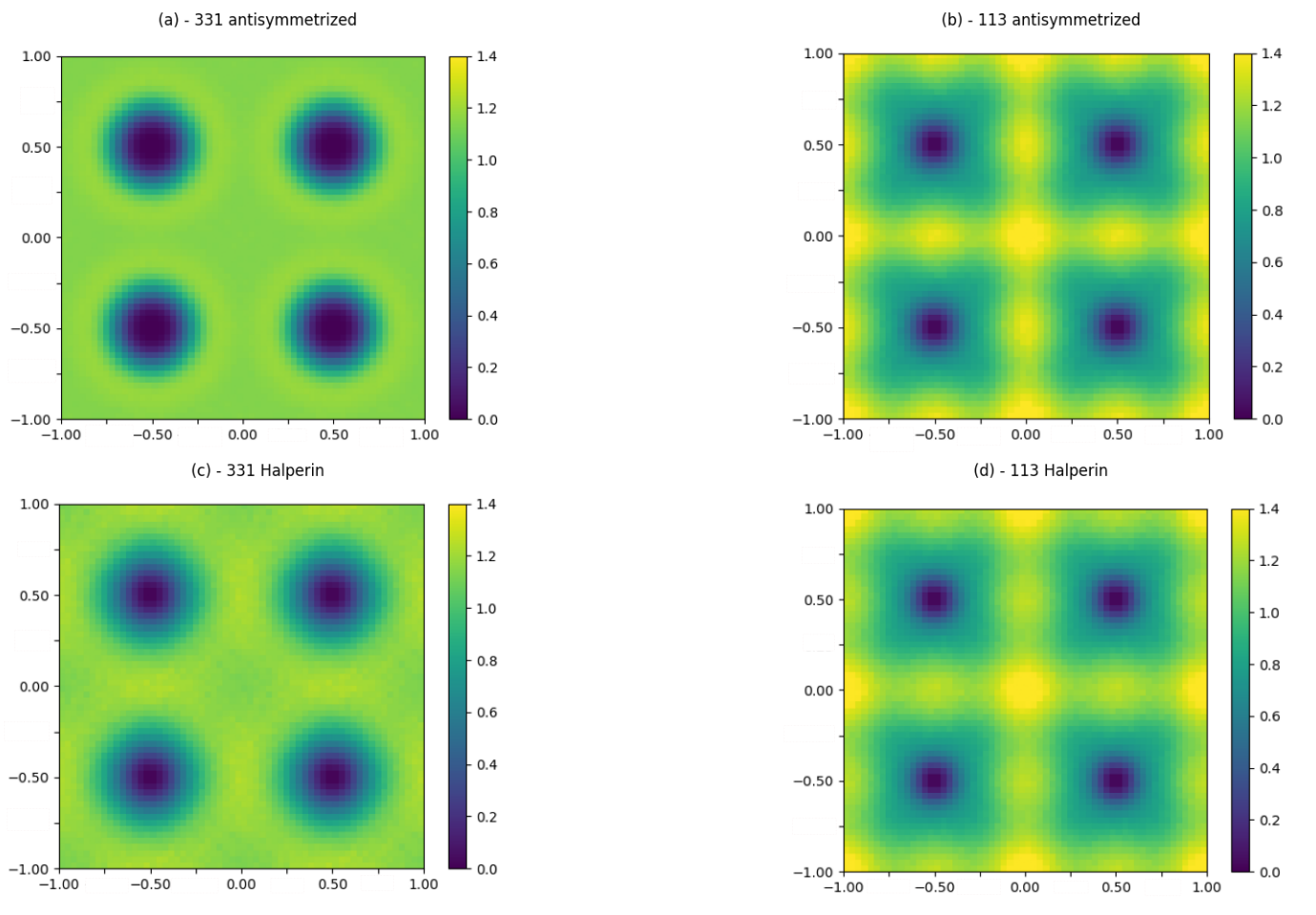}
    \caption{Two-point functions of fQHE for 331 (a) and 113 (b) Halperin states and 331 (c) and 113 (d) anti-symmetrized state with $N=8$ particles on the square torus.
    The coordinates on the axes are the usual reduced ones $x$ and $y$.
    Rotational symmetry breaking of the ground state is evident for the two 113 fQHE state (b-d), in comparison with the two 331 cases (a-c).
    We represented the same plot four times on the square in order to highlight the breaking of rotational symmetry. Note that the self-exclusion points are at $z=\pm\frac12\pm'\ii\frac12$ instead of at $z=0$.
    }
    
    \label{Fig:2ptFunctions}
\end{figure*}

\subsubsection{Two-point functions}
Finally, we also show the two-point function for the above-mentioned states.
Since we are particularly interested in how the 113 states may be sensitive to the torus geometry, we show the full 2D two-point function $g(z)$ rather than the integrated 1D version often used in literature.

In Figures~\ref{Fig:2ptFunctions}, for $N=8$ particles, we show the full two-point functions of the 113, 331, $\A$-113 and $\A$-331 states.
For pedagogic purposes, we show an extended version with the origin placed at $z=\pm\frac12\pm'\ii\frac12$.
It should be clear that both 331 and $\A$-331 show clearly rotationally symmetric exclusion holes (depleted regions), followed by an almost homogeneous region.

For the 113 and $\A$-113 states, the story is very different, since the correlation holes are no longer round, but take on the square shape of the torus itself. Furthermore, there is an over-density of particles, at the ``polar opposite side'' of the torus ($z=0$ in the image) as well as along the axes connecting a particle and its closest images.
At this state, we cannot conclusively say that we are observing the signature of a self-separating liquid, but it is remarkable the striking similarity between the 113 and $\A$-113 states, compared to how different they are from the normal FQH liquids as 331 and $\A$-331 states.

We have not explored the two-point functions as we change $\tau$, but speculate that the two-point functions of the $\A$-113 states should respond to geometric changes much more strongly than the 331 family.

\section{Discussion and conclusions}
\label{Discussion and Conclusions}
In this work, we have generalized the method that allows us to write fQHE wave functions on the torus, introduced by Haldane for the Laughlin state, to multi-variable fQHE states defined by generic integer-valued K-matrices; in particular, we have explicitly constructed trial wave functions of states with $K=(113)$ in their Halperin and anti-symmetrized configurations, and we validated our method with the state with $K=(331)$.
We numerically showed by computing the Hall viscosity that the putative $ A$-phase~\cite {ddm23} cannot be regarded as proper topological states.
We extended the use of clustering diagnostics, initially defined on the sphere, to the torus to substantiate our claim.
We find that these states are not fQHE states, as indicated by their geometry-dependent Hall viscosity.
We link the Hall viscosity dependence from the torus characteristic $\tau$ to the cluster formation in the space distribution of the particles in the proposed wave functions.
This picture agrees with the one presented in Ref.~\cite{sb24}, extending it to topological observables as the Hall viscosity.

By demonstrating the stable Hall viscosity for the Girvin-Jach extension of Laughlin states, we rule out the hypothesis that the $\tau$-dependence is solely due to the introduction of anti-holomorphic pieces in the CoM or to the study of states not at the LLL.
It can then be understood that the $\A$-113 family of states represents some compressible phase of matter that cannot reproduce the topological properties of usual fQHE states.
We expect our claim to be valid for the whole class of putative fQHE states with non-positive definite K-matrix~\cite{ddm23} since the $K=(113)$ case is the one that shows the smallest repulsion between different particle species in the Halperin construction and does not vanish after anti-symmetrization.

From a different perspective, in the context of the bulk-boundary correspondence, negative eigenvalues of the K-matrix encode the presence of counter-propagating modes at the edges of the sample.
It is then clear that ground-state wave functions of fQHE states with counter-propagating modes cannot be constructed from the short-distance behaviour ruled by the K-matrix itself.
Including anti-holomorphic pieces, as already shown in Ref.~\cite{hhsv17}, is key to stabilizing those states, at the cost of introducing higher LL components into the wave function.
This work shows that the topological information in the K-matrix must, in the case of counter-propagating (or negative eigenvalue) K-matrices, be augmented with short-distance details to produce stabilized fractional quantum Hall wave functions.

\begin{acknowledgments}
    We thank I. Gornyi, A. Mirlin and S. Simon for very useful discussion.
    This work is part of the D-ITP consortium, a program of the Dutch Research Council (NWO) that is funded by the Dutch Ministry of Education, Culture and Science (OCW).
    M.F. acknowledge the research program “Materials for the Quantum Age” (QuMat) for financial support. This program (registration number 024.005.006) is part of the Gravitation program financed by the Dutch Ministry of Education, Culture and Science (OCW).
\end{acknowledgments}

\bibliography{references}

@article{f17,
    author = {Fremling, Mikael},
    title = {Success and failure of the plasma analogy for Laughlin states on a torus},
    journal = {J. Phys. A: Math. Theor.},
    volume = {50},
    number = {015201},
    year = {2017}
}

@article{abgs98,
  title = {K-matrices for non-Abelian quantum Hall states},
  author = {Ardonne, Eddy and Bouwknegt, Peter and Guruswamy, Sathya and Schoutens, Kareljan},
  journal = {Phys. Rev. B},
  volume = {61},
  issue = {15},
  pages = {10298--10302},
  numpages = {0},
  year = {2000},
  month = {Apr},
  publisher = {American Physical Society},
  doi = {10.1103/PhysRevB.61.10298},
  url = {https://link.aps.org/doi/10.1103/PhysRevB.61.10298}
}

@Article{asz95,
  Title                    = {Viscosity of Quantum {H}all Fluids},
  Author                   = {Avron, J. E. and Seiler, R. and Zograf, P. G.},
  Journal                  = {Phys. Rev. Lett.},
  Year                     = {1995},
  Month                    = {Jul},
  Pages                    = {697--700},
  Volume                   = {75},
  Doi                      = {10.1103/PhysRevLett.75.697},
  Issue                    = {4},
  Numpages                 = {0},
  Publisher                = {American Physical Society},
  Url                      = {http://link.aps.org/doi/10.1103/PhysRevLett.75.697}
}

@article{pfj20,
  title = {Hall viscosity of composite fermions},
  author = {Pu, Songyang and Fremling, Mikael and Jain, J. K.},
  journal = {Phys. Rev. Res.},
  volume = {2},
  issue = {1},
  pages = {013139},
  numpages = {16},
  year = {2020},
  month = {Feb},
  publisher = {American Physical Society},
  doi = {10.1103/PhysRevResearch.2.013139},
  url = {https://link.aps.org/doi/10.1103/PhysRevResearch.2.013139}
}

@article{bons20,
  title = {Local Two-Body Parent Hamiltonians for the Entire Jain Sequence},
  author = {Bandyopadhyay, Sumanta and Ortiz, Gerardo and Nussinov, Zohar and Seidel, Alexander},
  journal = {Phys. Rev. Lett.},
  volume = {124},
  issue = {19},
  pages = {196803},
  numpages = {6},
  year = {2020},
  month = {May},
  publisher = {American Physical Society},
  doi = {10.1103/PhysRevLett.124.196803},
  url = {https://link.aps.org/doi/10.1103/PhysRevLett.124.196803}
}

@article{ddm23,
  title = {Anomalous Reentrant $5/2$ Quantum Hall Phase at Moderate Landau-Level-Mixing Strength},
  author = {Das, Sudipto and Das, Sahana and Mandal, Sudhansu S.},
  journal = {Phys. Rev. Lett.},
  volume = {131},
  issue = {5},
  pages = {056202},
  numpages = {6},
  year = {2023},
  month = {Aug},
  publisher = {American Physical Society},
  doi = {10.1103/PhysRevLett.131.056202},
  url = {https://link.aps.org/doi/10.1103/PhysRevLett.131.056202}
}

@article{ddm23reply,
  title = {Das et al. Reply:},
  author = {Das, Sudipto and Das, Sahana and Mandal, Sudhansu S.},
  journal = {Phys. Rev. Lett.},
  volume = {132},
  issue = {2},
  pages = {029602},
  numpages = {1},
  year = {2024},
  month = {Jan},
  publisher = {American Physical Society},
  doi = {10.1103/PhysRevLett.132.029602},
  url = {https://link.aps.org/doi/10.1103/PhysRevLett.132.029602}
}

@Article{f13,
  Title                    = {Coherent state wave functions on a torus with a constant magnetic field},
  Author                   = {Fremling, M.},
  Journal                  = {J. Phys. A: Math. Theor},
  Year                     = {2013},

  Month                    = {Jun},
  Pages                    = {275302},
  Volume                   = {46},

  Doi                      = {10.1088/1751-8113/46/27/275302},
  Eprint                   = {ar{X}iv:1302.6471},
  Issue                    = {27},
  Numpages                 = {20},
  Publisher                = {IOP Science},
  Url                      = {http://iopscience.iop.org/1751-8121/46/27/275302/}
}

@Article{ffms16,
  Title                    = {Energy projection and modified {L}aughlin states},
  Author                   = {Fremling, M. and Fulsebakke, J. and Moran, N. and Slingerland, J. K.},
  Journal                  = {Phys. Rev. B},
  Year                     = {2016},

  Month                    = {Jun},
  Pages                    = {235149},
  Volume                   = {93},

  Doi                      = {10.1103/PhysRevB.93.235149},
  Eprint                   = {ar{X}iv:1601.06736},
  Issue                    = {23},
  Numpages                 = {15},
  Publisher                = {American Physical Society},
  Url                      = {http://link.aps.org/doi/10.1103/PhysRevB.93.235149}
}

@Article{fhs14,
  Title                    = {Hall viscosity of hierarchical quantum {H}all states},
  Author                   = {Fremling, M. and Hansson, T. H. and Suorsa, J.},
  Journal                  = {Phys. Rev. B},
  Year                     = {2014},

  Month                    = {Mar},
  Pages                    = {125303},
  Volume                   = {89},

  Doi                      = {10.1103/PhysRevB.89.125303},
  Eprint                   = {ar{X}iv:1312.6038},
  Issue                    = {12},
  Numpages                 = {25},
  Publisher                = {American Physical Society},
  Url                      = {http://link.aps.org/doi/10.1103/PhysRevB.89.125303}
}

@Article{gj84,
  Title                    = {Formalism for the quantum {H}all effect: {H}ilbert space of analytic functions},
  Author                   = {Girvin, S. M. and Jach, Terrence},
  Journal                  = {Phys. Rev. B},
  Year                     = {1984},
  Month                    = {May},
  Pages                    = {5617--5625},
  Volume                   = {29},
  Doi                      = {10.1103/PhysRevB.29.5617},
  Issue                    = {10},
  Numpages                 = {0},
  Publisher                = {American Physical Society},
  Url                      = {http://link.aps.org/doi/10.1103/PhysRevB.29.5617}
}

@Article{gmp85,
  Title                    = {Magneto-roton theory of collective excitations in the fractional quantum {H}all effect},
  Author                   = {Girvin, S. M. and MacDonald, A. H. and Platzman, P. M.},
  Journal                  = {Phys. Rev. B},
  Year                     = {1986},
  Month                    = {Feb},
  Pages                    = {2481--2494},
  Volume                   = {33},
  Doi                      = {10.1103/PhysRevB.33.2481},
  Issue                    = {4},
  Publisher                = {American Physical Society},
  Url                      = {http://link.aps.org/doi/10.1103/PhysRevB.33.2481}
}

@Article{gmp85_2,
  Title                    = {Collective-Excitation Gap in the Fractional Quantum Hall Effect},
  Author                   = {Girvin, S. M. and MacDonald, A. H. and Platzman, P. M.},
  Journal                  = {Phys. Rev. Lett.},
  Year                     = {1985},

  Month                    = {Feb},
  Pages                    = {581--583},
  Volume                   = {54},

  Doi                      = {10.1103/PhysRevLett.54.581},
  Issue                    = {6},
  Numpages                 = {0},
  Publisher                = {American Physical Society},
  Url                      = {http://link.aps.org/doi/10.1103/PhysRevLett.54.581}
}

@article{wz92,
  title = {Classification of Abelian quantum Hall states and matrix formulation of topological fluids},
  author = {Wen, X. G. and Zee, A.},
  journal = {Phys. Rev. B},
  volume = {46},
  issue = {4},
  pages = {2290--2301},
  numpages = {0},
  year = {1992},
  month = {Jul},
  publisher = {American Physical Society},
  doi = {10.1103/PhysRevB.46.2290},
  url = {https://link.aps.org/doi/10.1103/PhysRevB.46.2290}
}

@Article{h83,
  Title                    = {Fractional Quantization of the {H}all Effect: A Hierarchy of Incompressible Quantum Fluid States},
  Author                   = {Haldane, F. D. M.},
  Journal                  = {Phys. Rev. Lett.},
  Year                     = {1983},

  Month                    = {Aug},
  Pages                    = {605--608},
  Volume                   = {51},

  Doi                      = {10.1103/PhysRevLett.51.605},
  Issue                    = {7},
  Numpages                 = {0},
  Publisher                = {American Physical Society},
  Url                      = {http://link.aps.org/doi/10.1103/PhysRevLett.51.605}
}

@article{h83_3,
    author = {Halperin, B.},
    title = {Theory of the quantized Hall conductance},
    journal = {Helvetica Physica Acta},
    year = {1983}
}

@Article{hr85,
  Title                    = {Periodic {L}aughlin-{J}astrow wave functions for the fractional quantized {H}all effect},
  Author                   = {Haldane, F. D. M. and Rezayi, E. H.},
  Journal                  = {Phys. Rev. B},
  Year                     = {1985},

  Month                    = {Feb},
  Pages                    = {2529--2531},
  Volume                   = {31},

  Doi                      = {10.1103/PhysRevB.31.2529},
  Issue                    = {4},
  Publisher                = {American Physical Society},
  Url                      = {http://link.aps.org/doi/10.1103/PhysRevB.31.2529}
}

@Article{hhrv09,
  Title                    = {Conformal Field Theory Approach to {A}belian and Non-{A}belian Quantum {H}all Quasielectrons},
  Author                   = {Hansson, T. H. and Hermanns, M. and Regnault, N. and Viefers, S.},
  Journal                  = {Phys. Rev. Lett.},
  Year                     = {2009},

  Month                    = {Apr},
  Pages                    = {166805},
  Volume                   = {102},

  Doi                      = {10.1103/PhysRevLett.102.166805},
  Eprint                   = {ar{X}iv:0810.0636},
  Issue                    = {16},
  Numpages                 = {4},
  Publisher                = {American Physical Society},
  Url                      = {http://link.aps.org/doi/10.1103/PhysRevLett.102.166805}
}

@Article{hhv09,
  Title                    = {Quantum {H}all quasielectron operators in conformal field theory},
  Author                   = {Hansson, T. H. and Hermanns, M. and Viefers, S.},
  Journal                  = {Phys. Rev. B},
  Year                     = {2009},

  Month                    = {Oct},
  Pages                    = {165330},
  Volume                   = {80},

  Doi                      = {10.1103/PhysRevB.80.165330},
  Eprint                   = {ar{X}iv:0903.0937},
  Issue                    = {16},
  Numpages                 = {22},
  Publisher                = {American Physical Society},
  Url                      = {http://link.aps.org/doi/10.1103/PhysRevB.80.165330}
}

@Article{hsbhk08,
  Title                    = {Quantum {H}all wave functions on the torus},
  Author                   = {Hermanns, M. and Suorsa, J. and Bergholtz, E. J. and Hansson, T. H. and Karlhede, A.},
  Journal                  = {Phys. Rev. B},
  Year                     = {2008},

  Month                    = {Mar},
  Pages                    = {125321},
  Volume                   = {77},

  Doi                      = {10.1103/PhysRevB.77.125321},
  Eprint                   = {ar{X}iv:0711.4684},
  Issue                    = {12},
  Numpages                 = {16},
  Publisher                = {American Physical Society},
  Url                      = {http://link.aps.org/doi/10.1103/PhysRevB.77.125321}
}

@Article{jk97,
  Title                    = {Composite Fermions in the Hilbert Space of the Lowest Electronic {L}andau Level},
  Author                   = {J. K. Jain and R. K. Kamilla},
  Journal                  = {Int. J. Mod. Phys. B},
  Year                     = {1997},
  Pages                    = {2621},
  Volume                   = {11},

  Doi                      = {10.1142/S0217979297001301},
  Eprint                   = {ar{X}iv:cond-mat/9704031},
  Issue                    = {22},
  Publisher                = {World Scientific},
  Url                      = {http://www.worldscientific.com/doi/abs/10.1142/S0217979297001301}
}

@Article{jk97_2,
  Title                    = {Quantitative study of large composite-fermion systems},
  Author                   = {Jain, J. K. and Kamilla, R. K.},
  Journal                  = {Phys. Rev. B},
  Year                     = {1997},

  Month                    = {Feb},
  Pages                    = {R4895--R4898},
  Volume                   = {55},

  Doi                      = {10.1103/PhysRevB.55.R4895},
  Issue                    = {8},
  Numpages                 = {0},
  Publisher                = {American Physical Society},
  Url                      = {http://link.aps.org/doi/10.1103/PhysRevB.55.R4895}
}

@article{kssj23,
  title = {Candidate local parent Hamiltonian for the 3/7 fractional quantum Hall effect},
  author = {Kudo, Koji and Sharma, A. and Sreejith, G. J. and Jain, J. K.},
  journal = {Phys. Rev. B},
  volume = {108},
  issue = {8},
  pages = {085130},
  numpages = {10},
  year = {2023},
  month = {Aug},
  publisher = {American Physical Society},
  doi = {10.1103/PhysRevB.108.085130},
  url = {https://link.aps.org/doi/10.1103/PhysRevB.108.085130}
}

@article{ntw85,
  title = {Quantized Hall conductance as a topological invariant},
  author = {Niu, Qian and Thouless, D. J. and Wu, Yong-Shi},
  journal = {Phys. Rev. B},
  volume = {31},
  issue = {6},
  pages = {3372--3377},
  numpages = {0},
  year = {1985},
  month = {Mar},
  publisher = {American Physical Society},
  doi = {10.1103/PhysRevB.31.3372},
  url = {https://link.aps.org/doi/10.1103/PhysRevB.31.3372}
}

@article{wn90,
  title = {Ground-state degeneracy of the fractional quantum Hall states in the presence of a random potential and on high-genus Riemann surfaces},
  author = {Wen, X. G. and Niu, Q.},
  journal = {Phys. Rev. B},
  volume = {41},
  issue = {13},
  pages = {9377--9396},
  numpages = {0},
  year = {1990},
  month = {May},
  publisher = {American Physical Society},
  doi = {10.1103/PhysRevB.41.9377},
  url = {https://link.aps.org/doi/10.1103/PhysRevB.41.9377}
}

@Article{k03,
  Title                    = {Fault-tolerant quantum computation by anyons },
  Author                   = {A.Yu. Kitaev},
  Journal                  = {Annals of Physics },
  Year                     = {2003},
  Number                   = {1},
  Pages                    = {2 - 30},
  Volume                   = {303},

  Doi                      = {http://dx.doi.org/10.1016/S0003-4916(02)00018-0},
  Eprint                   = {ar{X}iv:quant-ph/9707021},
  ISSN                     = {0003-4916},
  Url                      = {http://www.sciencedirect.com/science/article/pii/S0003491602000180}
}

@Article{l83,
  Title                    = {Anomalous Quantum {H}all Effect: An Incompressible Quantum Fluid with Fractionally Charged Excitations},
  Author                   = {Laughlin, R. B.},
  Journal                  = {Phys. Rev. Lett.},
  Year                     = {1983},

  Month                    = {May},
  Pages                    = {1395--1398},
  Volume                   = {50},

  Doi                      = {10.1103/PhysRevLett.50.1395},
  Issue                    = {18},
  Publisher                = {American Physical Society},
  Url                      = {http://link.aps.org/doi/10.1103/PhysRevLett.50.1395}
}

@Article{lrnf07,
  Title                    = {Particle-Hole Symmetry and the $\ensuremath{\nu}=\frac{5}{2}$ Quantum {H}all State},
  Author                   = {Lee, Sung-Sik and Ryu, Shinsei and Nayak, Chetan and Fisher, Matthew P. A.},
  Journal                  = {Phys. Rev. Lett.},
  Year                     = {2007},

  Month                    = {Dec},
  Pages                    = {236807},
  Volume                   = {99},

  Doi                      = {10.1103/PhysRevLett.99.236807},
  Issue                    = {23},
  Numpages                 = {4},
  Publisher                = {American Physical Society},
  Url                      = {http://link.aps.org/doi/10.1103/PhysRevLett.99.236807}
}

@Article{lhr07,
  Title                    = {Particle-Hole Symmetry and the {P}faffian State},
  Author                   = {Levin, Michael and Halperin, Bertrand I. and Rosenow, Bernd},
  Journal                  = {Phys. Rev. Lett.},
  Year                     = {2007},

  Month                    = {Dec},
  Pages                    = {236806},
  Volume                   = {99},

  Doi                      = {10.1103/PhysRevLett.99.236806},
  Issue                    = {23},
  Numpages                 = {4},
  Publisher                = {American Physical Society},
  Url                      = {http://link.aps.org/doi/10.1103/PhysRevLett.99.236806}
}

@Article{mr91,
  Title                    = {Nonabelions in the fractional quantum {H}all effect },
  Author                   = {Gregory Moore and Nicholas Read},
  Journal                  = {Nuclear Physics B },
  Year                     = {1991},
  Number                   = {2-3},
  Pages                    = {362 - 396},
  Volume                   = {360},

  Doi                      = {http://dx.doi.org/10.1016/0550-3213(91)90407-O},
  ISSN                     = {0550-3213},
  Url                      = {http://www.sciencedirect.com/science/article/pii/055032139190407O}
}

@Article{nssfd08,
  Title                    = {Non-Abelian anyons and topological quantum computation},
  Author                   = {Nayak, Chetan and Simon, Steven H. and Stern, Ady and Freedman, Michael and Das Sarma, Sankar},
  Journal                  = {Rev. Mod. Phys.},
  Year                     = {2008},

  Month                    = {Sep},
  Pages                    = {1083--1159},
  Volume                   = {80},

  Doi                      = {10.1103/RevModPhys.80.1083},
  Eprint                   = {ar{X}iv:0707.1889},
  Issue                    = {3},
  Numpages                 = {0},
  Publisher                = {American Physical Society},
  Url                      = {http://link.aps.org/doi/10.1103/RevModPhys.80.1083}
}

@Article{qjm89,
    Title                  = {{P}hases of the multiple quantum well in a strong magnetic field: {P}ossibility of irrational charge},
    Author                 = {Qiu, X. and Joint, R. and MacDonald, A.},
    Journal                = {Phys. Rev. B},
    Year                   = {1989},

    Month                  = {Dec},
    Pages                  = {11943},
    Volume                 = {40},

    Doi                    = {10.1103/PhysRevB.40.11943},
    Issue                  = {17},
    Publisher              = {American Physical Society},
    Url                    = {https://doi.org/10.1103/PhysRevB.40.11943}
    }

@Article{r09,
  Title                    = {Non-Abelian adiabatic statistics and Hall viscosity in quantum {H}all states and ${p}_{x}+i{p}_{y}$ paired superfluids},
  Author                   = {Read, N.},
  Journal                  = {Phys. Rev. B},
  Year                     = {2009},

  Month                    = {Jan},
  Pages                    = {045308},
  Volume                   = {79},

  Doi                      = {10.1103/PhysRevB.79.045308},
  Eprint                   = {ar{X}iv:0805.2507},
  Issue                    = {4},
  Numpages                 = {41},
  Publisher                = {American Physical Society},
  Url                      = {http://link.aps.org/doi/10.1103/PhysRevB.79.045308}
}

@article{rr08,
  title = {Hall viscosity, orbital spin, and geometry: Paired superfluids and quantum Hall systems},
  author = {Read, N. and Rezayi, E. H.},
  journal = {Phys. Rev. B},
  volume = {84},
  issue = {8},
  pages = {085316},
  numpages = {27},
  year = {2011},
  month = {Aug},
  publisher = {American Physical Society},
  doi = {10.1103/PhysRevB.84.085316},
  url = {https://link.aps.org/doi/10.1103/PhysRevB.84.085316}
}

@article{sb24,
  title = {Phase separation in the putative fractional quantum Hall $\mathcal{A}$ phases},
  author = {Simon, Steven H. and Balram, Ajit C.},
  journal = {Phys. Rev. B},
  volume = {111},
  issue = {4},
  pages = {045102},
  numpages = {12},
  year = {2025},
  month = {Jan},
  publisher = {American Physical Society},
  doi = {10.1103/PhysRevB.111.045102},
  url = {https://link.aps.org/doi/10.1103/PhysRevB.111.045102}
}

@article{s23,
  title = {Comment on ``Anomalous Reentrant $5/2$ Quantum Hall Phase at Moderate Landau-Level-Mixing Strength''},
  author = {Simon, Steven H.},
  journal = {Phys. Rev. Lett.},
  volume = {132},
  issue = {2},
  pages = {029601},
  numpages = {2},
  year = {2024},
  month = {Jan},
  publisher = {American Physical Society},
  doi = {10.1103/PhysRevLett.132.029601},
  url = {https://link.aps.org/doi/10.1103/PhysRevLett.132.029601}
}

@article{sfjj18,
  title = {Search for exact local Hamiltonians for general fractional quantum Hall states},
  author = {Sreejith, G. J. and Fremling, M. and Jeon, Gun Sang and Jain, J. K.},
  journal = {Phys. Rev. B},
  volume = {98},
  issue = {23},
  pages = {235139},
  numpages = {18},
  year = {2018},
  month = {Dec},
  publisher = {American Physical Society},
  doi = {10.1103/PhysRevB.98.235139},
  url = {https://link.aps.org/doi/10.1103/PhysRevB.98.235139}
}

@Article{svh11_2,
  Title                    = {A General Approach to Quantum {H}all Hierarchies},
  Author                   = {Suorsa, J. and Viefers, S. and Hansson, T. H.},
  Journal                  = {New J. Phys.},
  Year                     = {2011},
  Pages                    = {075006},
  Volume                   = {13},

  Doi                      = {10.1088/1367-2630/13/7/075006},
  Eprint                   = {ar{X}iv:1011.5365},
  Url                      = {http://iopscience.iop.org/1367-2630/13/7/075006}
}

@article{hhsv17,
  title = {Quantum Hall physics: Hierarchies and conformal field theory techniques},
  author = {Hansson, T. H. and Hermanns, M. and Simon, S. H. and Viefers, S. F.},
  journal = {Rev. Mod. Phys.},
  volume = {89},
  issue = {2},
  pages = {025005},
  numpages = {61},
  year = {2017},
  month = {May},
  publisher = {American Physical Society},
  doi = {10.1103/RevModPhys.89.025005},
  url = {https://link.aps.org/doi/10.1103/RevModPhys.89.025005}
}

@Article{w92,
  author  = {Wen, X.G.},
  title   = {Theory of the Edge Excitations in {FQH} effects},
  journal = {Int. J. Mod. Phys.},
  year    = {1992},
  volume  = {B6},
  pages   = {1711},
  doi     = {10.1142/S0217979292000840}
}

@article{h11,
  title = {Geometrical Description of the Fractional Quantum Hall Effect},
  author = {Haldane, F. D. M.},
  journal = {Phys. Rev. Lett.},
  volume = {107},
  issue = {11},
  pages = {116801},
  numpages = {5},
  year = {2011},
  month = {Sep},
  publisher = {American Physical Society},
  doi = {10.1103/PhysRevLett.107.116801},
  url = {https://link.aps.org/doi/10.1103/PhysRevLett.107.116801}
}

@article{zaletel2013,
  title={Topological Characterization of Fractional Quantum Hall Ground States from Microscopic Hamiltonians},
  author={Zaletel, Michael P and Mong, Roger SK and Pollmann, Frank},
  journal={Physical review letters},
  volume={110},
  number={23},
  pages={236801},
  year={2013},
  publisher={APS}
}

@book{Fradkin91,
    author = {E. Fradkin},
    title = {Field Theories of Condensed Matter Physics},
    publisher = {Cambridge University Press},
    year = {1991}
}

@book{Mumford07,
    author = {Mumford, D.},
    title = {Tata Lectures on Theta I},
    publisher = {Birkhäuser Boston, MA},
    year = {2007}
}

@book{ChaikinLubensky95,
    author = {Chaikin, P.M. and Lubensky, T.C.},
    title = {Principles of condensed matter physics},
    publisher = {Cambridge University Press},
    year = {1995}
}

@article{k85,
    author = {Kohmoto, M.},
    title = {Topological invariant and the quantization of the Hall conductance},
    journal = {Ann. Phys.},
    volume = {160},
    issue = {343},
    year = {1985}
}

@article{l30,
    author = {L. Landau},
    title = {Diamagnetismus der Metalle},
    journal = {Z. Physik},
    volume = {64},
    pages = {629–637},
    year = {1930},
    url = {https://doi.org/10.1007/BF01397213}
}

@article{z64,
  title = {Magnetic Translation Group},
  author = {Zak, J.},
  journal = {Phys. Rev.},
  volume = {134},
  issue = {6A},
  pages = {A1602--A1606},
  numpages = {0},
  year = {1964},
  month = {Jun},
  publisher = {American Physical Society},
  doi = {10.1103/PhysRev.134.A1602},
  url = {https://link.aps.org/doi/10.1103/PhysRev.134.A1602}
}

@article{s15,
  title = {Is the Composite Fermion a Dirac Particle?},
  author = {Son, Dam Thanh},
  journal = {Phys. Rev. X},
  volume = {5},
  issue = {3},
  pages = {031027},
  numpages = {14},
  year = {2015},
  month = {Sep},
  publisher = {American Physical Society},
  doi = {10.1103/PhysRevX.5.031027},
  url = {https://link.aps.org/doi/10.1103/PhysRevX.5.031027}
}

@article{hbfgwtjz25,
    author = {Huang, K. and Balram, A.C. and Fu, H. and Guo, C. and Watanabe, K. and Taniguchi, T. and Jain, J.K. and Zhu, J.},
    title = {Hetero-Orbital Two-Component Fractional Quantum Hall States in Bilayer Graphene},
    journal = {arXiv},
    year = {2025},
    url = {https://arxiv.org/abs/2506.14188}
}

@article{Haldane1985Laughlin,
  title = {Periodic Laughlin-Jastrow wave functions for the fractional quantized Hall effect},
  author = {Haldane, F. D. M. and Rezayi, E. H.},
  journal = {Phys. Rev. B},
  volume = {31},
  issue = {4},
  pages = {2529--2531},
  numpages = {2},
  year = {1985},
  month = {Feb},
  publisher = {American Physical Society},
  doi = {10.1103/PhysRevB.31.2529},
  url = {https://link.aps.org/doi/10.1103/PhysRevB.31.2529}
}

@book{DiFrancesco:1997nk,
    author = "Di Francesco, P. and Mathieu, P. and Senechal, D.",
    title = "{Conformal Field Theory}",
    doi = "10.1007/978-1-4612-2256-9",
    isbn = "978-0-387-94785-3, 978-1-4612-7475-9",
    publisher = "Springer-Verlag",
    address = "New York",
    series = "Graduate Texts in Contemporary Physics",
    year = "1997"
}

@article{ch1958,
    author = {Cahn and Hilliard},
    title = {Free Energy of a Nonuniform System. I. Interfacial Free Energy},
    journal = {J. Chem. Phys.},
    volume = {28},
    pages = {258--267},
    numpages = {10},
    year = {1958},
    month = {Feb},
    publisher = {AIP Publishing},
    doi = {https://doi.org/10.1063/1.1744102}
}

\appendix

\section{Effective field theory of quantum Hall effect}
\label{App: Effective Field Theory of Quantum Hall effect}
In this section we introduce the effective $(2+1)$D Chern-Simons field theory that describes the low-energy properties of fQHE by the means of emergent fields $\Phi^{a}_\mu$ coupled each other via the K-matrix and with a background electromagnetic field $A_\mu$ as
\begin{equation}
    \label{CSaction}
    S[\phi,A] = \frac{1}{4\pi}\int \dd^3 x \epsilon^{\mu\nu\rho}\left(K^{ab}\phi^a_\mu\partial_\nu\phi^b_\rho + 2t_c A_\mu\partial_\nu\phi^c_\rho\right)~,
\end{equation}
where the $\vec{t}$ vector specifies which linear combination of emergent fields has to be considered the proper electronic current, given that
\begin{equation}
    \label{CScurrent}
    J^\mu = \frac{\delta S}{\delta A_\mu} = \frac{1}{2\pi}\epsilon^{\mu\nu\rho}t_c\partial_\nu\phi^c_\rho~.
\end{equation}
The K-matrix and the vector $\vec{t}$ encode then all the topological properties of the theory (filling fraction $\nu$, quasi-hole charge vector $\vec{e}^*$ and statistics between quasi-hole $\alpha^{ab}$)
\begin{eqnarray}
    \label{CSTopo}
    &\nu = \vec{t}^TK^{-1}\vec{t}~, \\
    &\vec{e}^* = K^{-1}\vec{t}~, \\
    &\alpha^{ab} = \left(K^{-1}\right)^{ab}~.
\end{eqnarray}
The degeneracy of the ground states on a torus is given by $\det(K)$.

Both K-matrix and vector $\vec{t}$ are not unique: they can both be mapped into others by a linear transformation $\vec{t}\to S\vec{t}$ and $K\to SKS^{-1}$, induced by the invertible matrix $S$.
This means physically that the electronic basis $\vec{t}=\vec{1}$ is not a special description of the strongly-interacting problem; on the contrary, the most special basis is the one that diagonalizes the K-matrix and it basically splits the multivariate problem as an algebraic sum of single-variable problems.
In the case of positive-definite K-matrix, all the eigenvalues are positive by definition and the effective action \eqref{CSaction} produces a well-defined partition function $Z$ which is the product of all the single partition functions; if this requirement is not met, then this is not possible since the partition function that describes the problem is ill-defined, since it is the ratio between different single variables partition functions.

\section{Theta functions}
\label{App: Theta functions}
In this appendix we explicitly define the building blocks of our wave functions: the theta functions in $d$ space dimensions; we especially focus on the two main cases, \ie $d=1$ and $d=2$ dimensions.
In the first section we introduce their series representation and the remaining two we list their quasi-periodic and modular properties.
As a side, we include the Dedekind eta function in our discussion.

\subsection{Series representation}
\label{App: Series representation}
In order to define theta functions with $d$-dimensional complex vector $\vec{Z}$ as an argument, we introduce a $d\times d$-dimensional complex-valued matrix $\mathbf{\Omega}$ with positive-definite imaginary part which plays the same role as $\tau$ in the mono-dimensional case.
We employ then the following identification
\begin{eqnarray}
    \label{GenModCs}
    \vec{Z} \sim \vec{Z} + \hat{e}_j ~, \\
    \vec{Z} \sim \vec{Z} + \mathbf{\Omega}\hat{e}_j ~,
\end{eqnarray}
after a translation of a unit in the $j$-direction, labeled by the unit vector $\hat{e}_j$, and we impose periodic boundary conditions on the real direction while quasi-periodic on the complex one.
By doing so and expanding using Fourier transform, we can prove that theta functions are holomorphic on the complex plane (singularities are placed at infinity) and they can be represented as infinite series
\begin{equation}
    \label{thetaMD}
    \vartheta\begin{bmatrix} \vec{\alpha} \\ \vec{\beta}\end{bmatrix}(\vec{Z}|\mathbf{\Omega}) = \sum_{\vec{m}\in\mathbb{Z}^d}e^{\ii\pi(\vec{m}+\vec{\alpha})^T\mathbf{\Omega}(\vec{m}+\vec{\alpha}) + 2\ii\pi(\vec{m}+\vec{\alpha})^T(\vec{Z}+\vec{\beta})}.
\end{equation}
Where the label $T$ stands for matrix transposition.
We directly stated the formula with characteristic vectors $\vec{\alpha}$ and $\vec{\beta}$.
Few comments are necessary: this representation is meaningful if and only if the imaginary part of $\mathbf{\Omega}$ is positive definite and the first argument of the exponential dominates the regions with large values of the components of $\vec{m}$; if it converges, it does it uniformly and absolutely, meaning that this can estimate potentially any point on the complex plane.
The positive-definiteness of the imaginary part of $\mathbf{\Omega}$ is related to the orientability of the manifold: when this is absent, it is impossible to define a smooth covering of the multi-dimensional torus.
Moreover, one can prove that these functions vanish at the point
\begin{equation}
    \label{thetaZeros}
    \vec{Z}_0 = \left(\vec{\beta}+\frac{1}{2}\vec{1}\right) + \mathbf{\Omega}\left(\vec{\alpha}+\frac{1}{2}\vec{1}\right)~,
\end{equation}
where we employed the shorthand $\vec{1} = (1,\dots,1)$; given periodicity and quasi-periodicity this point is not unique, but there is an infinite number of them.

We now specify the generic formula to the $d=1$ case, where the matrix $\mathbf{\Omega}$ is the scalar $\tau$
\begin{equation}
    \label{theta1D}
    \vartheta\begin{bmatrix} \alpha \\ \beta\end{bmatrix}(Z|\tau) = \sum_{m=-\infty}^{+\infty}e^{\ii\pi\tau(m+\alpha)^2+ 2\ii\pi(m+\alpha)(Z+\beta)},
\end{equation}
and the multi-dimensional expression takes a very simple form.
A particular role is played by those theta functions with characteristics $\alpha=\beta=1/2$, since their zero is placed at $Z=0$ and in the neighborhood $Z\to 0$ they can be approximated as $Z$, \ie they do not depend on the torus characteristic $\tau$ and they are linear.

For completeness, we also write down the detailed formula for the case $d=2$
\begin{align}
    \label{theta2D}
    &\vartheta\begin{bmatrix} \vec{\alpha} \\ \vec{\beta}\end{bmatrix}(\vec{Z}|\mathbf{\Omega}) = \sum_{m_1=-\infty}^{+\infty}\sum_{m_2=-\infty}^{+\infty} \\
    &e^{\ii\pi\left[\mathbf{\Omega}_{11}(m_1+\alpha_1)^2 + 2\mathbf{\Omega}_{12}(m_1+\alpha_1)(m_2+\alpha_2) + \mathbf{\Omega}_{22}(m_2+\alpha_2)^2\right]} \nonumber\\
    &\times~e^{2\ii\pi\left[(m_1+\alpha_1)(Z_1+\beta_1)+(m_2+\alpha_2)(Z_2+\beta_2)\right]}~.\nonumber
\end{align}

Finally we introduce the Dedekind eta function as an infinite product
\begin{equation}
    \label{DedekindEta}
    \eta(\tau) = e^\frac{\ii\pi\tau}{12}\prod_{m=1}^\infty(1-e^{2\ii\pi m\tau})~.
\end{equation}

\subsection{Quasi-periodic properties}
\label{App: Quasi-periodic properties}
Employing the series representation \eqref{thetaMD} it is possible to prove the following periodic and quasi-periodic behavior
\begin{eqnarray}
    \label{P-QPMD}
    &\vartheta\begin{bmatrix} \vec{\alpha} \\ \vec{\beta}\end{bmatrix}(\vec{Z}+\vec{l}_1 + \mathbf{\Omega}\vec{l}_2|\mathbf{\Omega}) = e^{2\ii\pi\chi(\vec{l}_1,\vec{l}_2)}\vartheta\begin{bmatrix} \vec{\alpha} \\ \vec{\beta}\end{bmatrix}(\vec{Z}|\mathbf{\Omega}), \\
    &\chi(\vec{l}_1,\vec{l}_2) = \vec{\alpha}^T\vec{l}_1 - \vec{\beta}^T\vec{l}_2 - \frac{1}{2}\vec{l}_2^T\mathbf{\Omega}\vec{l}_2 - \vec{l}_2^T\vec{Z},
\end{eqnarray}
by the means of (multi-dimensional) Poisson resummation formula for integer $\vec{l}_1$ and $\vec{l}_2$.
Please note that the term $\chi$ is complex, since it is linear in $\vec{Z}$ and $\mathbf{\Omega}$, although it turns into a simple phase when the translation is along the real direction (periodicity)
\begin{equation}
    \label{PMD}
    \vartheta\begin{bmatrix} \vec{\alpha} \\ \vec{\beta}\end{bmatrix}(\vec{Z}+\vec{l}_1|\mathbf{\Omega}) = e^{2\ii\pi\vec{\alpha}^T\vec{l}_1}\vartheta\begin{bmatrix} \vec{\alpha} \\ \vec{\beta}\end{bmatrix}(\vec{Z}|\mathbf{\Omega})~.
\end{equation}
The norm of the theta function changes when translations are applied in any direction $\mathbf{\Omega}\hat{e}_j$ (quasiperiodicity)
\begin{equation}
    \label{QPMD}
    \vartheta\begin{bmatrix} \vec{\alpha} \\ \vec{\beta}\end{bmatrix}(\vec{Z}+ \mathbf{\Omega}\vec{l}_2|\mathbf{\Omega}) = e^{-\ii\pi\vec{l}_2^T\mathbf{\Omega}\vec{l}_2}e^{-2\ii\pi\vec{l}_2^T(\vec{Z}+\vec{\beta})}\vartheta\begin{bmatrix} \vec{\alpha} \\ \vec{\beta}\end{bmatrix}(\vec{Z}|\mathbf{\Omega}),
\end{equation}
which is necessary in order to define an holomorphic function on the whole complex plane.

We again state the explicit realizations for $d=1$, for the periodicity on the real direction
\begin{equation}
    \label{P1D}
    \elliptic\alpha\beta{Z + l_1}{\tau} = e^{2\ii\pi\alpha l_1}\elliptic\alpha\beta Z\tau~,
\end{equation}
and the quasi-periodicity along the $\tau$ direction 
\begin{equation}
    \label{QP1D}
    \elliptic\alpha\beta{Z + l_2\tau}\tau = e^{-\ii\pi\tau l_2^2}e^{-2\ii\pi l_2(Z+\beta)}\elliptic\alpha\beta Z\tau~.
\end{equation}

\subsection{Modular properties}
\label{App: Modular properties}
Being that theta functions are defined on the torus, they must also map themselves into their class when the torus get reparametrized, \ie they must transform in a covariant way under the action of the modular group.
In the $d$-dimensional setting it is customary to describe the modular transformations as the action of a symplectic matrix $\mathbf{\Gamma}$ of dimensions $2d\times2d$, formed by $4$ other $d\times d$ dimensional matrices $\mathbf{A}$, $\mathbf{B}$, $\mathbf{C}$ and $\mathbf{D}$ with integer elements
\begin{equation}
    \label{SymplMatr}
    \mathbf{\Gamma} = \begin{pmatrix} \mathbf{A} & \mathbf{B} \\ \mathbf{C} & \mathbf{D}\end{pmatrix}.
\end{equation}
The elements of the matrix $\mathbf{\Gamma}$ act in the following way on the arguments $\vec{Z}$ and $\mathbf{\Omega}$ of the multi-dimensional theta function
\begin{eqnarray}
    \label{MTZOmega}
    \vec{Z}' = \left[(\mathbf{C}\mathbf{\Omega}+\mathbf{D})^{-1}\right]^T\vec{Z}, \\
    \mathbf{\Omega}' = (\mathbf{A}\mathbf{\Omega}+\mathbf{B})(\mathbf{C}\mathbf{\Omega}+\mathbf{D})^{-1},
\end{eqnarray}
while the characteristics
\begin{eqnarray}
    \label{MTchar}
    \vec{\alpha}' = \mathbf{D}\vec{\alpha} - \mathbf{C}\vec{\beta} + \frac{1}{2}\vec{\mathrm{diag}}(\mathbf{CD}^T),\\
    \vec{\beta}' = -\mathbf{B}\vec{\alpha} + \mathbf{A}\vec{\beta} + \frac{1}{2}\vec{\mathrm{diag}}(\mathbf{AB}^T),
\end{eqnarray}
with the shorthand $\vec{\mathrm{diag}}(\mathbf{A})$ for the vector containing all the diagonal elements of the matrix $\mathbf{A}$.
The action on the theta function itself reads

    \begin{align}
        \label{MTMD}
    \vartheta\begin{bmatrix} \vec{\alpha}' \\ \vec{\beta}'\end{bmatrix}(\vec{Z}'|\mathbf{\Omega}') = \kappa(\vec{\alpha},\vec{\beta},\mathbf{\Gamma})\sqrt{\det(\mathbf{C}\mathbf{\Omega}+\mathbf{D})}\nonumber\\
    \times e^{\ii\pi\vec{Z}^T(\mathbf{C}\mathbf{\Omega}+\mathbf{D})^{-1}\mathbf{C}\vec{Z}}\vartheta\begin{bmatrix} \vec{\alpha} \\ \vec{\beta}\end{bmatrix}(\vec{Z}|\mathbf{\Omega}),
    \end{align}

where $\kappa$ is a simple phase depending on the characteristics and the modular transformation parameters, but it does not have an easy closed form in the multi-dimensional realization.
Such formula allows to define the main three possible types of modular transformations:
\begin{enumerate}
    \item A simple change of basis in the $d$-dimensional space
    \begin{equation}
        \mathbf{\Gamma} = \begin{pmatrix} \mathbf{A} & \mathbf{0} \\ \mathbf{0} & \left(\mathbf{A}^{-1}\right)^T\end{pmatrix},
    \end{equation}
    which simply rescales the multi-dimensional theta function.

    \item A shift in the real direction, \ie $T$ transformation in the multi-dimensional space
    \begin{equation}
        \mathbf{\Gamma} = \begin{pmatrix} \mathbf{I} & \mathbf{B} \\ \mathbf{0} & \mathbf{I}\end{pmatrix},
    \end{equation}
    which keeps the theta function invariant, apart from a shift in the characteristics $\vec{\beta}$, equivalent to a shift in the argument $\vec{Z}$.

    \item The $S$ transformation
    \begin{equation}
        \mathbf{\Gamma} = \begin{pmatrix} \mathbf{0} & -\mathbf{I} \\ \mathbf{I} & \mathbf{0}\end{pmatrix},
    \end{equation}
    that introduces a non-trivial exponential factor in the transformed theta function
    \begin{equation}
        \vartheta\begin{bmatrix} \vec{\alpha}' \\ \vec{\beta}'\end{bmatrix}(\vec{Z}'|\mathbf{\Omega}') = \sqrt{\det(-\ii\mathbf{\Omega})}e^{\ii\pi\vec{Z}^T\mathbf{\Omega}^{-1}\vec{Z}}\vartheta\begin{bmatrix} \vec{\alpha} \\ \vec{\beta}\end{bmatrix}(\vec{Z}|\mathbf{\Omega}).
    \end{equation}
\end{enumerate}

In the $d=1$ dimensional setting the possible transformations remain three, but the first one is usually neglected as long as it is a simple rescaling of the variable $Z$ by an integer number.
The other two are usually written in the following form
\begin{eqnarray}
    \label{MT1Dtheta}
    \vartheta\begin{bmatrix} \alpha \\ \beta\end{bmatrix}(Z|\tau + l) = e^{-\ii\pi\alpha(1+\alpha)l}\vartheta\begin{bmatrix} \alpha \\ \alpha l + \frac{l}{2} + \beta\end{bmatrix}(Z|\tau),\nonumber \\
    \vartheta\begin{bmatrix} \alpha \\ \beta\end{bmatrix}\left(Z|-\frac{1}{\tau}\right) = e^{2\ii\alpha\beta}\sqrt{-\ii\tau}e^{\ii\pi\tau Z^2}\vartheta\begin{bmatrix} \beta \\ -\alpha\end{bmatrix}(\tau Z|\tau), \nonumber
\end{eqnarray}
where the $\kappa$ phase can be uniquely specified.

The Dedekind eta has peculiar properties under modular transformations; as a matter of facts, the following relations hold
\begin{eqnarray}
    \label{DedekindModTrans}
    \eta\left(-\frac{1}{\tau}\right) = \sqrt{\ii\tau} \eta(\tau)~, \\
    \eta(\tau+1) = e^\frac{\ii\pi}{12}\eta(\tau)~.
\end{eqnarray}

\section{Derivation of center-of-mass pieces}
\label{App: Derivation of center-of-mass pieces}
In this section we explicitly derive the CoM pieces of the torus wave functions for different candidate fQHE states.

\subsection{Laughlin state}
\label{App: Laughlin state}
As we introduced in the main text, Laughlin states are characterized by the Jastrow factor \eqref{JFL1}.
One can easily prove, using the properties of monodimensional theta functions, that
\begin{eqnarray}
    \label{JFLP-QP}
    &J(z_1+1,\dots|\tau) = (-1)^{N-1} J(z_1,\dots|\tau)~,\\
    &J(z_1+\tau,\dots|\tau) =\nonumber\\
    &(-1)^{N-1}\left[\prod_{j=1}^N e^{-\ii\pi\tau q + 2\ii\pi\tau q(z_1-z_j)}\right]J(z_1,\dots|\tau),
\end{eqnarray}
where the dots represents all other coordinates, up to the $N$-th one; the same transformation is valid for any other $i$-th coordinate.
Note that imaginary part of $z_i$-dependent term is canceled by the transformation of the Gaussian piece \eqref{GaussianPBC}.
This implies that the CoM has to cancel the following contribution $e^{\ii\pi q\tau  + 2\ii\pi qZ}$, namely
\begin{equation}
    \label{CoMLQP}
    \mathcal{F}_{\alpha,\beta}(Z+\tau|\tau) = e^{-\ii\pi q\tau - 2\ii\pi qZ}\mathcal{F}_{\alpha,\beta}(Z|\tau)~,
\end{equation}
up to a phase, determined by the characteristics $\alpha$ and $\beta$.
The candidate CoM piece must also be periodic along the real direction, hence a theta function is the only holomorphic solution to the problem
\begin{equation}
    \label{CoMLDerivation}
    \mathcal{F}_{\alpha,\beta}(Z|\tau) = \vartheta\begin{bmatrix} \alpha \\ q\beta\end{bmatrix}(qZ|q\tau)~,
\end{equation}
but we still need to fix the values of the characteristics, to check the modular properties and the ground state degeneracy.

The first task can be solved by studying the CoM translations $T_1$ and $T_2^q$: the Jastrow factor, being dependent only from relative coordinates, is invariant under both their action, hence under $T_1$ the whole wave function only changes by the factor $e^{2\ii\pi q\alpha}$.
Under $T_2^q$, the transformation is more structured, since the Gaussian piece develops the prefactor $e^{\ii\pi\tau q (2Y + 1)}$, where $Y = \sum_{i=1}^N y_i$ and the CoM transform as \eqref{QP1D}, introducing a prefactor $e^{-\ii\pi\tau q  - 2\ii\pi qZ )}$ that cancels the Gaussian one, leaving the usual $X$-dependent phase that is removed by a gauge transformation.
In addition to this part there is also the contribution of characteristics, that is proportional to $e^{2\ii\pi\beta}$; given that we are strictly imposing periodic boundary conditions, we have to put $\beta$ vanishing or proportional to an integer; on the other hand, $\alpha$ can take all the fractional values $s/q$, where $s$ is an integer smaller than $q$.

The requirement of modular invariance is fundamental to determine the normalization of the whole wave function that depends on the torus parameter $\tau$.
We start from the analysis of the $T$-transformation: the Gaussian piece gets a pure phase \eqref{GaussianModTrans} that is compensated by a gauge transformation and the other pieces provide a constant phase that can be removed afterwards without relying on gauge transformations; the interesting bit in this case comes from the change in the boundary conditions after a CoM translation since the parameter $\beta$, that was set to zero before in order to accommodate for periodic boundary conditions, but it is now shifted by a factor of $q/2$, which is half-odd by construction and leads to anti-periodic boundary conditions.

In the $S$-transformation we have similar result, since the Jastrow factor develops a multiplicative factor that cancels the one introduced by the Gaussian piece \eqref{GaussianModTrans} and another one $e^{\ii\pi q Z^2/\tau}$ that has to be removed by the CoM piece.
The CoM needs a bit more of effort to derive its $T$-transformation, since the usual relations for $d=1$ dimensional theta functions \eqref{MT1Dtheta} are not valid due to the presence of the constant $q$ in the arguments; however, we can employ Poisson resummation and obtain the following expression
\begin{equation}
    \label{CoMMTTTrans}
    \vartheta\begin{bmatrix} 0 \\ -s/q \end{bmatrix}\left(q\frac{\tau Z}{|\tau|}|-\frac{q}{\tau}\right) = \sqrt{\ii\tau}e^{-\ii\pi q Z^2/\tau}\vartheta\begin{bmatrix} s/q \\ 0\end{bmatrix}\left(qZ|q\tau\right),
\end{equation}
that removes the factor introduced by the Jastrow part and ensures the correct modular invariance, apart from the presence of the factor $\sqrt{\ii\tau}$; such contribution can be removed with the help of the inverse of Dedekind eta function, given \eqref{DedekindModTrans}.
We retrieve in the end the result \eqref{CoML1} we stated in the main text.

Finally we can show that the degeneracy of the ground states is exactly the number of zeroes of the CoM piece; given that the CoM vanishes at
\begin{equation}
    qZ_0 = \frac{1}{2} + \frac{1}{2}\tau ~,
\end{equation}
in the fundamental region of the torus, then it has $q$ different zeroes in it, as predicted by Wen-Zee classification
\begin{equation}
    \label{CoMLZeroes}
    Z_{0,j} = \frac{1+2j}{2q} + \frac{q+2s}{2q}\tau ~,
\end{equation}
where $j$ runs from $0$ to $q-1$.

\subsection{Laughlin state with Girvin-Jach Jastrow part}
\label{App: Laughlin state with Girvin-Jach Jastrow part}
The Girvin-Jach construction \eqref{EQ:GJ-WF} introduces anti-holomorphic pieces in the wave function and hence it can be qualitatively understood not as the ground state of a fQHE with filling fraction $\nu=1/q$, but as one of the excited states.
The CoM in this case can be derived in the same fashion as it was found in the previous section, but there is a simpler way.
We can construct the full wave function in this case by taking the Laughlin wave functions for states with $q+k$ and $k$ inverse filling fraction, \ie $\psi_{q+k}(\vec{z},\tau)$ and $\psi_{k}(\vec{z},\tau)$.
Then the product
\begin{equation}
    \label{GJWF}
    \psi^{GJ}_{q,k}(\vec{z},\tau) = \psi_{q+k}(\vec{z},\tau)\left(\psi_k(\vec{z},\tau)\right)^*
\end{equation}
transforms correctly under translations of periods along the real and $\tau$ directions on the torus.
It also presents the same Jastrow factor \eqref{EQ:GJ-WF}, the Gaussian piece is correctly defined as in \eqref{QHEwavefunction} and the CoM can be read straightforwardly from this expression as
\begin{eqnarray}
    \label{CoML2}
    &\mathcal{F}_s(Z|\tau)=\left|\frac{1}{\eta(\tau)}\right|^2\vartheta\begin{bmatrix} s/(q+k) \nonumber\\ 0\end{bmatrix}((q+k)Z|(q+k)\tau)\\&\times\left(\vartheta\begin{bmatrix} s/k \\ 0\end{bmatrix}(kZ|k\tau)\right)^*.
\end{eqnarray}
One can show that this is not the ground state of a fQHE state since the number of zeroes is larger than $q$, \ie there are more magnetic fluxes attached to the particles than the minimal number.

\subsection{The 331 state}
\label{App: 331 state}
In the case of a multi-species Jastrow factor, determined by a positive-definite K-matrix, the CoM piece is more elaborate.
As a matter of facts, the Jastrow piece transforms as
\begin{eqnarray}
    \label{JF331P-QP}
    &J(z_1+1,\dots|\tau) = (-1)^{N-1} J(z_1,\dots|\tau)~,\\
    &J(z_1+\tau,\dots|\tau) =\nonumber\\
    &\left[e^{\ii\pi\tau (K_{11} - N_\phi) - 2\ii\pi\tau (N_\phi z_1 - (K\vec{Z})_i)}\right]J(z_1,\dots|\tau),
\end{eqnarray}
and after removing terms due to the transformation of the Gaussian piece \eqref{QHEwavefunction}, the CoM has to transform in the following way
\begin{equation}
    \label{CoM331QP}
    \mathcal{F}_{\vec{\alpha},\vec{\beta}}(Z+\tau\hat{e}_a|\tau) = e^{-\ii\pi K_{aa}\tau - 2\ii\pi K_{ab}Z_b}\mathcal{F}_{\alpha,\beta}(Z|\tau)~,
\end{equation}
when a translation in the group of particles $a=1,2$ takes place; we write the product between the K-matrix and the CoM coordinate vector $\vec{Z}$ as $K_{ab}Z_b$.
Given that the CoM has to be periodic with reference to translations in the real direction, it can only be written in terms of a $d=2$ dimensional theta function and it reads
\begin{equation}
    \label{CoM331Derivation}
    \mathcal{F}_{\vec{\alpha},\vec{\beta}}(\vec{Z}|\tau)=\vartheta\begin{bmatrix} \vec{\alpha} \\ \vec{\beta}\end{bmatrix}(K\vec{Z}|\tau K)~.
\end{equation}
In this case we cannot straightforwardly show it enjoys invariance under the action of CoM translations: to get it, we first observe that \eqref{CoM331Derivation} is already invariant under the action of $T_1$, since $K\vec{v}=\vec{1}$, and then we only need to have an expression that is invariant under the action of $T_2^q$; it can be achieved by simply taking the sum $\mathcal{F}_{\vec{0},\vec{0}} + \mathcal{F}_{2\vec{v},\vec{0}}$.
Such linear combination preserves the modular invariance, while the degeneracy of the ground state exactly reproduces its number of zeroes predicted by the determinant of the K-matrix.

It can be shown that the same result can be retrieved by employing $d=1$ dimensional theta functions.
The method we dubbed monodromy matching starts by making the following ansatz for the building block of the CoM piece
\begin{eqnarray}
    \label{MMAnsatz}
    &\G_{\vec{\alpha},\vec{\beta}}(\vec{Z}|\tau)= \left(\vartheta\begin{bmatrix} \alpha_1 \\ \beta_1\end{bmatrix}(Z_1|\tau) \right)^{d_1}\left(\vartheta\begin{bmatrix} \alpha_2 \\ \beta_2\end{bmatrix}(Z_2|\tau) \right)^{d_2} \nonumber\\
    &\times\left(\vartheta\begin{bmatrix} \alpha_1 + \alpha_2 \\ \beta_1 + \beta_2\end{bmatrix}(Z_1+Z_2|\tau) \right)^{d_0}~,
\end{eqnarray}
and by requiring the same periodicity and CoM translation properties one derives the linear combination of $\G$ that reproduces the equivalent version of the $d=2$ dimensional theta function.
In the specific case of $K = (3 3 1)$ states, the CoM building block obtained this way reads the following
\begin{eqnarray}
    \label{MM331BB}
    &\G_{\vec{\alpha},\vec{\beta}}(\vec{Z}|\tau) = \left(\vartheta\begin{bmatrix}
			\alpha_1 \\
			\beta_1
		\end{bmatrix}(Z_1|\tau)\right)^2\left(\vartheta\begin{bmatrix}
		\alpha_2 \\
		\beta_2
	\end{bmatrix}(Z_2|\tau)\right)^2\nonumber \\
    &\times\vartheta\begin{bmatrix}
	\alpha_1 +\alpha_2 \\
	\beta_1 + \beta_2
	\end{bmatrix}(Z_1+Z_2|\tau)~,
\end{eqnarray}
\ie $d_1=d_2=2$ and $d_0=1$.
The linear combination that provides the correct invariant CoM piece has then the following form
\begin{eqnarray}
		\label{MMLC331}
		&C(\tau)\left[\mathcal{F}_{\vec{0},\vec{0}}(\vec{Z}|\tau) + \mathcal{F}_{2\vec{v},\vec{0}}(\vec{Z}|\tau)\right] =  \\
        &\G_{\vec{0},\vec{0}}(\vec{Z}|\tau) + \G_{2\vec{v},\vec{0}}(\vec{Z}|\tau) + 
		\G_{\vec{0},\vec{v}}(\vec{Z}|\tau) + \G_{2\vec{v},\vec{v}}(\vec{Z}|\tau) \nonumber\\
		&+\G_{\vec{0},2\vec{v}}(\vec{Z}|\tau) + \G_{2\vec{v},2\vec{v}}(\vec{Z}|\tau) +
		\G_{\vec{0},3\vec{v}}(\vec{Z}|\tau) + \G_{2\vec{v},3\vec{v}}(\vec{Z}|\tau) ,\nonumber
\end{eqnarray}
by the means of the vector $\vec{v}=(1/4)~\vec{1}$, up to a constant prefactor $C$, which is shown in Appendix \ref{App: Relation between different center-of-mass pieces}.

\subsection{113 state}
\label{App: 113 state}
When the state is described by a non positive-definite K-matrix it is not possible to write the CoM piece in terms of $d=2$ dimensional theta function.
However, as described in the main text, the monodromy matching ansatz can be still employed and it provides negative values for the exponents $d_1$ and $d_2$, which we set equal for simplicity.
Such negative values can be interpreted either as fractions of theta functions or as complex conjugates of the latter; in the second case, boundary conditions has to be revisited since transformation rules of the building blocks are changed
\begin{eqnarray}
    \label{MMQP}
    &\G_{\vec{\alpha},\vec{\beta}}(\vec{Z}+\hat{e}_j\tau|\tau) =\\ &e^{\ii\pi(d_0\tau-|d_1|\bar{\tau})}e^{2\ii\pi(d_0Z_1 - |d_1|\bar{Z_1} + |d_1|Z_2)}
    \G_{\vec{\alpha},\vec{\beta}}(\vec{Z}|\tau)\nonumber,
\end{eqnarray}
requiring the cancellation of the exponentially decaying term $e^{\ii\pi(d_0\tau-|d_1|\bar{\tau})(1-2Y_1)}$ by the introduction of an extra Gaussian piece
\begin{equation}
    \label{MMGaussCorr}
    G_{AH}(\vec{Y}|\tau) = e^{2\ii\pi\left[(d_0\tau-|d_1|\bar{\tau})(Y_1^2+Y_2^2)\right]}~.
\end{equation}
These results can be specified for the proposed case by imposing $d_1=d_2=K_{11}-K_{12}$ and $d_0=K_{12}$.

\section{Relation between different center-of-mass pieces}
\label{App: Relation between different center-of-mass pieces}

In this section we show that the two expressions
\begin{subequations}
    \begin{align}
    &\mathcal{F}_{\vec{\alpha},\vec{\beta}}(\vec{Z}|\tau) = \vartheta\begin{bmatrix}
        \vec{\alpha} \\
        \vec{\beta}
    \end{bmatrix}(K\vec{Z}|\tau K)~,\label{331F}  \\
    &\G_{\vec{\alpha},\vec{\beta}}(\vec{Z}|\tau) = \left(\vartheta\begin{bmatrix}
        \alpha_1 \\
        \beta_1
    \end{bmatrix}(Z_1|\tau)\right)^2\left(\vartheta\begin{bmatrix}
    \alpha_2 \\
    \beta_2
\end{bmatrix}(Z_2|\tau)\right)^2\nonumber \\ &\times\vartheta\begin{bmatrix}
\alpha_1 +\alpha_2 \\
\beta_1 + \beta_2
\end{bmatrix}(Z_1+Z_2|\tau)~,\label{331G}
\end{align}
\end{subequations}
provide the same CoM piece for the 331 state, up to a prefactor depending only from the torus parameter $\tau$ and the K-matrix.
The first theta function is defined for multidimensional complex variable while the ones contained in the second expression are the monodimensional ones.
The CoM coordinates $\vec{Z} = (Z_1,Z_2)$ and the characteristic vectors $\vec{\alpha}=(\alpha_1,\alpha_2)$ and $\vec{\beta} = (\beta_1,\beta_2)$ are the same for both expressions.
Another useful quantity is the vector $\vec{v} = (1/4)~\vec{1}$.
The expressions for the CoM piece match when linear combinations of these building blocks are taken into account
\begin{align}
    \label{MainEq}
    &C(\tau)\left[\mathcal{F}_{\vec{0},\vec{0}}(\vec{Z}|\tau) + \mathcal{F}_{2\vec{v},\vec{0}}(\vec{Z}|\tau)\right] = \\ &\G_{\vec{0},\vec{0}}(\vec{Z}|\tau) + \G_{2\vec{v},\vec{0}}(\vec{Z}|\tau) + 
    \G_{\vec{0},\vec{v}}(\vec{Z}|\tau) + \G_{2\vec{v},\vec{v}}(\vec{Z}|\tau) \nonumber\\
    &+\G_{\vec{0},2\vec{v}}(\vec{Z}|\tau) + \G_{2\vec{v},2\vec{v}}(\vec{Z}|\tau) +
    \G_{\vec{0},3\vec{v}}(\vec{Z}|\tau) + \G_{2\vec{v},3\vec{v}}(\vec{Z}|\tau) .\nonumber
\end{align}
Both the expressions transform in the same way for translations of the $j$-th particle's group of a unit vector $\vec{e}_j$ in the $x$ and $\tau$ directions; this can be easily checked by using \eqref{P-QPMD}.
More importantly, both the expressions must be eigenvalues of the CoM translation operators $T_1$ and $T_2^2$, which amount to translations of $\vec{v}$ and $2\tau\vec{v}$.
For the first row of \eqref{MainEq}, it is important to see that $K\vec{v} = \vec{1}$, reducing the translation along the real axis to simple shifts of a period, \ie $\mathcal{F}$ is already an eigenvalue of $T_1$; the translation along the $y$ direction is not of a fully period, but half of it, hence we need to consider the linear combination of the two contributions $\mathcal{F}_{\vec{0},\vec{0}}(\vec{Z}|\tau)$ and $\mathcal{F}_{\vec{0},2\vec{v}}(\vec{Z}|\tau)$ in order to fulfill periodicity for CoM translations.
Note that such linear combination is still an eigenvalue of $T_1$.
For what regards the second row, the function $\G$ is neither an eigenvalue of $T_1$ nor $T_2^2$, and we simply need to consider a linear combination of all the possible points in the space of $\vec{\alpha}$ and $\vec{\beta}$ inside the unit cell onto which the origin is mapped.
By definition the two expressions are holomorphic functions with the same periodicity, hence their ratio should not depend on the complex variable $z$.

We can further constraint the relation between the constant and the torus characteristic $\tau$ imposing modular invariance for both the sides in \eqref{MainEq}.
We only have to derive the action of the modular group on \eqref{331G}, since for \eqref{331F} it is already contained in Appendix~\ref{App: Modular properties}.
Employing the form of modular transformations of single monodimensional theta functions in \eqref{MTMD} we get, after a $T$ transformation
\begin{equation}
    \G_{\vec{\alpha},\vec{\beta}}(\vec{Z}|\tau) = e^{\ii\pi\vec{\alpha}^TK(\vec{1}+\vec{\alpha})}\G_{\vec{\alpha},\vec{\alpha} + (1/2)\vec{1} + \vec{\beta}}(\vec{Z}|\tau)~,
\end{equation}
the same transformation that \eqref{331F} enjoys; hence $T$ transformations are not very informative in order to determine the dependence of $C$ from $\tau$.
On the other hand, $S$ transformations act in the following way
\begin{equation}
    \G_{\vec{\alpha},\vec{\beta}}\left(\vec{Z}|-\frac{1}{\tau}\right) = e^{\ii\pi\vec{\alpha}^TK\vec{\beta}}(\sqrt{-\ii\tau})^5 e^{\ii\pi\tau\vec{Z}^TK\vec{Z}}\G_{\vec{\beta},-\vec{\alpha}}(\tau\vec{Z}|\tau)~,
\end{equation}
which contains the correct exponential of $\vec{Z}^TK\vec{Z}$ and form of ${G}_{\vec{\beta},-\vec{\alpha}}(\tau\vec{Z}|\tau)$, but the $\tau$ dependent part differs.
Hence, the prefactor $C(\tau)$ must transform as
\begin{equation}
    C\left(-\frac{1}{\tau}\right) = (\sqrt{-\ii\tau})^3 C(\tau)~,
\end{equation}
and it can be expressed at this point as
\begin{equation}
    C(\tau) = \mathcal{C} [\eta(\tau)]^3~,
\end{equation}
in terms of the Dedekind eta function.
The remaining unknown part $\mathcal{C}$ is a purely numerical factor that enters the normalization of the state and can always be rescaled without any loss of generality.

Finally, we would stress that we had not proved analytically, but just numerically, the strict identity between the two products of theta functions \eqref{331F} and \eqref{331G}; the technical reason behind it lies in the fact that decomposing the square of a theta function in the space of theta functions is not a trivial exercise~\cite{Mumford07}.
Our analytical derivation lies on the equivalence between pseudoperiodic holomorphic functions with the same period and the same transformation rules.

\section{CoM translations and linear dependence}\label{App:CoM-basis}
When considering fixing boundary conditions, it should be clear that there are usually more variables $\{\alpha\}$ and $\{\beta\}$ than there are equations to constrain them.
The total number or variables is $\sum_{\vec i}(d^U_{\vec i}+d^L_{\vec i})$ whereas the number of equations are $2n$.
For, \eg the 331 state, this leads to the relation equations $\beta_{1,1}+\beta_{1,2}+\beta_{12}=_1 \frac12$, and $\beta_{2,1}+\beta_{2,2}+\beta_{12}=_1 \frac12$  which leaves 3 out of 5 variables undetermined (and similarly for $\alpha$).

The example above raises the question of what the true degeneracy of these states is.
To answer this question is both easy and hard at the same time.
We can show numerically that the degeneracy of all the states considered matches with the expectation from the determinant of the K-matrix.
However, to prove this analytically is a non-trivial task, as we will explain below.

To set the stage, let us first consider the available quantum numbers related to the action of the reduced CoM operators $\tilde T_j$ (from equations \ref{eq:CoM-Tilde-Def} and \ref{eq:CoM-Tilde}) on $\G$.
One can show that applying $\tilde T_1$ on $\G(\vec Z)$ is equivalent to making in in \eqref{MMAnsatzGeneral} (or eqn. \eqref{MMAnsatzGeneral_II}) the substitution
\begin{equation}
  \label{eq:CoM_subs}
  \beta_{\vec i,k}\to \beta_{\vec i,k} + \sum_{j\in\vec i}\frac{N_j}{N_\phi} = \beta_{\vec i,k} + \nu_{\vec{i}},
\end{equation}
where for brevity we write $\nu_{\vec{i}}=\sum_{j\in\vec i}\nu_{j}$.
For $\tilde T_2$, the same holds but with $\beta$ replaced by $\alpha$.
For the 331 state this means, \eg that $\tilde T_1\G(\vec Z)$ gives $\beta_{1,k}\to\beta_{1,k}+\frac14$, $\beta_{2,k}\to\beta_{2,k}+\frac14$ and $\beta_{12}\to\beta_{12}+\frac12$ respectively.
From equation \eqref{eq:CoM_subs} we find that e.g. $\tilde T_j^4$ is an identity operation, since $\tilde T_j^4\G_{331}\propto\G_{331}$.
In general, $\tilde T_j^{q^\star} \sim 1$ where $q^\star$ is given by the minimal cycle that sends all $\beta_{\vec i,k}\to \beta_{\vec i,k} + \mathbb Z$.
In other words we seek the smallest $q^\star\in\mathbb Z^+$ such that $q^\star\nu_{\vec{i}} =_1 0$ for all $\vec i$.

Starting from a given reference state $\G$, we can now construct a basis for the CoM pieces as
\begin{equation}
  \label{eq:CoM-basis}
  \G^{(k_1,k_2)}\propto \sum_{j_1=1}^{q^\star}\sum_{j_2=1}^{q^\star/q} e^{\frac{2\pi}{q^\star}(j_1k_1+q j_2k_2)}\,T_1^{k_1}T_2^{qk_2}\G.
\end{equation}
which are simultaneously eigenstates of $T_1$ and $T_2^q$. We see that the number of states is $q^\star \times q^\star/q$.

For the Laughlin states, then $q^\star=q$, effectively giving a one-dimentional space $G_q^{(k_1,k_2)}\to G_q^{(k_1)}$, as expected for an abelian state.
For the 331 state, then $q=2$ and $q^\star=4$, which gives a basis of dimension $4\times2=8$, which is consistent with the degeneracy of the K-matrix.

We have further checked numerically that any $\G_{331}$, with arbitrary coefficient {$\alpha$} and {$\beta$}, could be completely expanded in the CoM basis \eqref{eq:CoM-basis}~\footnote{``any'' in the sense of all $\G_{331}$ we generated numerically, while still respecting the constraints coming from boundary conditions.}.
We expect that the phenomenon of over-parametrization will be true in general and that the monodromy matching will always produce a basis with dimension $|\det(K)|$. However, we have not attempted to prove it.
We suspect that what is going on is, in spirit, the same as the redundancies created by the trigonometric identities on the space of sines and cosines.
There, one can show that, \eg the function $\sin(\theta+\beta)$ can always be expressed as a linear combination of the functions $\sin(\theta)$ and $\cos(\theta)$ through the relation $\sin(\theta+\beta)=\sin(\theta)\cos(\beta)+\cos(\theta)\sin(\beta)$.
Something similar is surely happening for $\G$ but driven by relationships between theta functions, of which there are many.

\section{Boundary energy of self-separated plasma}
\label{App: Boundary energy of self-separated plasma}
Self-separated fluids induce clustering among their components and they are characterized by a surface tension between the two liquids.
The contribution to the energy given by the surface tension was introduced by Cahn and Hilliard~\cite{ch1958}, which reduces to the well-known phenomenological energy of a surface $A$~\cite{ChaikinLubensky95} given by the terms
\begin{equation}
    \label{HBoundary}
    H_B = H_\sigma + H_c + H_G.
\end{equation}
The equation contains the surface tension term
\begin{equation}
    H_\sigma = \int dA~\sigma(\vec{n}),
\end{equation}
dependent from the surface normal $\vec{n}$; the mean curvature one
\begin{equation}
    H_c = \kappa\int dA \left(\frac{1}{R_1}+\frac{1}{R_2} - \frac{2}{R_0}\right)^2,
\end{equation}
where $R_1$ and $R_2$ are the principal curvatures and $R_0$ represents a preferred curvature; and the Gaussian term
\begin{equation}
    H_G = \kappa_G\int dA \frac{1}{R_1R_2} = 4\pi(1-g),
\end{equation}
which depends only from the genus $g$ of the boundary surface for the Gauss-Bonnet theorem~\cite{ChaikinLubensky95}.

We aim to discuss the properties of the self-separated plasma at thermal equilibrium, and can make some simplifications.
For instance, we can assume that the Gaussian term does not play any role, $\kappa_G = 0$.
Since the system is defined on a flat manifold, we can also impose that $R_0=\infty$.
Finally, we can assume $\sigma(\vec{n})$ is an analytic function of the surface normal, and we can expand it for small deviations of the height $h$.
This leads to the following harmonic Hamiltonian
\begin{equation}
    H_B = \sigma A + \int dx [\gamma(\partial h)^2+\kappa(\partial^2 h)^2],
\end{equation}
where the interfacial stiffness $\gamma>0$ as well as the mean curvature contribution $\kappa>0$.
The Hamiltonian~\ref{HBoundary} is clearly gapless, since the energy spectrum allows for small momentum excitations at small energies.
Moreover, there is no notion of charge, and the deforming modes are completely neutral.

As a final comment, the fact that the two self-separated plasma phases are both in their ``liquid" regime (\ie their relative inverse filling fraction from the K-matrix is lower than the critical value $q_c$, usually $q_c\sim70$ for Laughlin states) allows for having an analytical for of the surface tension density $\sigma(\vec{n})$, which is not the case for solids~\cite{ChaikinLubensky95}.
This, in turn, implies that roughening is forbidden in these cases and that the surface modes are gapless.

\section{The divergence at $\tau=\i$}
\label{App: The divergence}
\begin{figure}[t]
    \centering
    \includegraphics[width=\columnwidth]{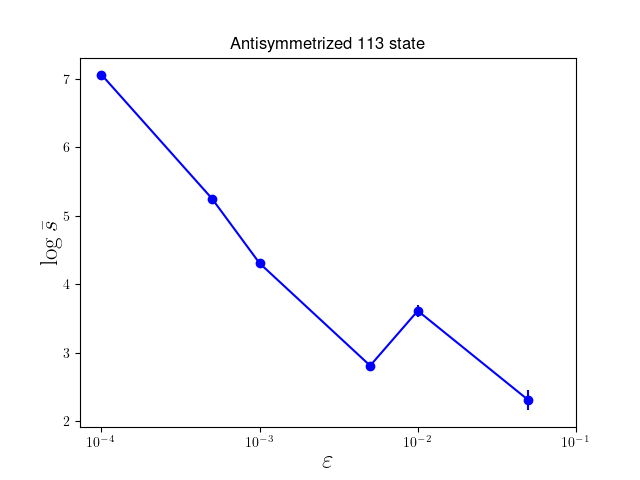}
    \caption{Hall viscosity encoded by the average orbital spin, $\sbar$, for the $\A$-113 state as a function of the side of the square, $\varepsilon$, for $N=6$ particles.
      The plot clearly shows divergent behaviour for $\varepsilon$ approaching zero.
      We conclude that the ``finiteness'' of the peak in Figure \ref{Fig:HallViscosityTau} in the main text is actually a finite-size effect of $\varepsilon$.}
    \label{Fig:HallViscosityPH}
\end{figure}

\begin{figure}[t]
    \centering
    \includegraphics[width=\columnwidth]{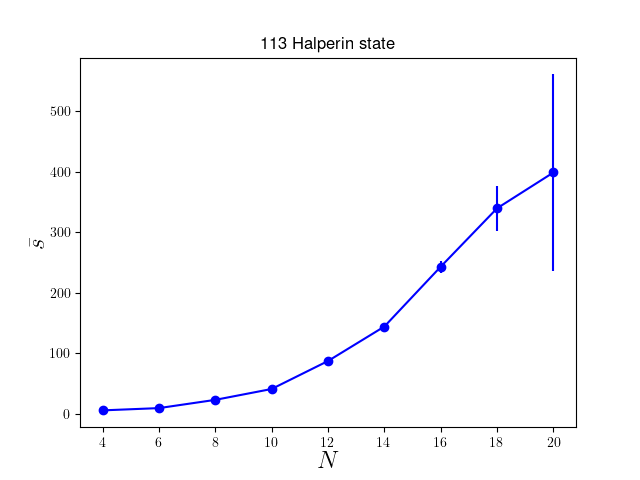}
    \caption{Hall viscosity encoded by average orbital spin for the 113 Halperin state. We fix the dimension of the square to be $\varepsilon = 0.001$ while we vary particle number $N$. Hall viscosity monotonously increases as a function of $N$ showing that the divergence gets more pronounced at larger system sizes.}
    \label{Fig:HallViscosityHal}
\end{figure}

In this section we take a closer look at the divergence in viscosity at $\tau=\i$.
As we can see in Figure~\ref{Fig:HallViscosityPH} for $N=6$ particles, the anti-symmetrized state diverges as $\varepsilon$ tends to smaller values, while in Figure~\ref{Fig:HallViscosityHal} we show that Halperin state's viscosity grows with the number of particles $N$, which means that this topological quantity is not defined in the thermodynamic limit.


\section{Geometric artifacts in cluster diagnostics}
\label{App: Geometric artifacts in cluster diagnostics}
In this section, we list the differences with the work~\cite{sb24} concerning the geometric effects when determining the clustering diagnostics on a square torus ($\tau_2 = 1$), instead of a sphere.
The main issue emerges from the different topology of the two surfaces: the first is the simplest genus-1 surface, the second is not.

We start by defining length on a torus as the shortest distance between two points, taking periodicity into account.
In practice, it amounts to the following formula for two points $z_1$ and $z_2$
\begin{equation}
  \label{TorusDist}
    d(z_1,z_2) = \left|\overline{\text{mod}}(x_d)_1 + \ii\cdot\overline{\text{mod}}(y_d)_1 \right|
\end{equation}
where we defined $z_d\equiv z_1 - z_2 = x_d + \ii y_d$. The function $\overline{\text{mod}}(x)_a=\text{mod}(x+\frac{a}2,a)-\frac{a}2$ is the shifted modulo function which bring $x$ back into the range $(-\frac{a}2,\frac{a}2]$, and effectively finds the closest ``image'' of $z_d$.
  The distance $d(z_1,z_2)$ enters the computation of the distance with the closest $k$-th particle in~\cite{sb24} and trivially removes the artifacts of having holonomies, namely that there are multiple topologically non-equivalent paths between two points.

We can also define the area of a circle of radius $r$ in the periodic complex plane, which is cropped by a square fundamental domain.
The area is represented by the piecewise function
\begin{equation}\label{TorusArea}
    A(r)=\left\{\begin{alignedat}{5}
        & \pi r^2, \ \ \ \text{if }~ r<\frac{1}{2},  \\
        & [\pi - 4\Delta(r)]r^2 , \ \ \ \text{if } ~\frac{1}{2}<r<\frac{1}{\sqrt{2}}, \\
        & 1, \ \ \ \text{if }~ r>1,
    \end{alignedat}\right.
\end{equation}
with the help of the auxiliary function
\begin{equation}
    \Delta(r) = \cos^{-1}\left(\frac{1}{2r}\right) - \frac{1}{2}\sqrt{\frac{1}{r^2} - \frac{1}{4r^4}}~.
\end{equation}
We employ the area to show the correct rescaled distances among the particles on the torus by exploiting the relation between the area that encloses $k$ different particles with the distance between the $k$-th closest particle and the particle at the center in the homogeneous case, with constant density $\rho$.
In comparison with the sphere, the area of a circle on the torus cannot be differentiated twice in the two points $r_1 = 1/2$ and $r_2 = 1/\sqrt{2}$, that leads to deformations on the simple relation $A(r_k)\rho = k$ which are taken into account when presenting our results.



\end{document}